\documentclass[twocolumn]{aastex62}
\usepackage{CJK}
\usepackage{hyperref}

\usepackage{ulem}

\newcommand{\ssst}{\scriptscriptstyle}
\newcommand{\E}[1]{\times 10^{#1}}

\newcommand{\RAdot}[4]{{#1}^{{\rm h}}{#2}^{{\rm m}}{#3}\fs{#4}}
\newcommand{\decldot}[4]{{#1}^{\circ}{#2}'{#3}\farcs{#4}}

\newcommand{\ps}{\,{\rm s}^{-1}}
\newcommand{\yr}{\,{\rm yr}}    
\newcommand{\Msun}{M_{\odot}}
\newcommand{\cm}{\,{\rm cm}}    
\newcommand{\km}{\,{\rm km}}
\newcommand{\kms}{$\km\ps$}

\newcommand{\kpc}{\,{\rm kpc}} 
\newcommand{\pc}{\,{\rm pc}}
\newcommand{\erg}{\,{\rm erg}}  
\newcommand{\K}{\,{\rm K}}

\newcommand{\um}{\,\mu\rm m}

\newcommand{\Rs}{R_{\rm s}}

\newcommand{\vs}{v_{s}}

\newcommand{\nHH}{n({\rm H}_{2})} 

\newcommand{\VLSR}{V_{\ssst\rm LSR}}

\newcommand{\Spitzer}{{\sl Spitzer}}

\newcommand{\XMMN}{{\sl XMM-Newton}}

\newcommand{\Chandra}{{\sl Chandra}}
\newcommand{\du}{d_{6.6}}

\newcommand{\snr}{G57.2$+$0.8}
\newcommand{\sgr}{SGR~1935$+$2154}
\newcommand{\twCO}{$^{12}$CO}   
\newcommand{\thCO}{$^{13}$CO}

\newcommand{\Jotz}{$J$=1--0}

\graphicspath{{./}{figures/}}

\received{May 7, 2020}
\accepted{October 5, 2020}

\begin{document}

\begin{CJK*}{UTF8}{bsmi}

\title{Revisiting the distance, environment and supernova properties of SNR G57.2+0.8 that hosts SGR 1935+2154}
\author{Ping Zhou}
\affil{Anton Pannekoek Institute for Astronomy, University of Amsterdam, Science Park 904, 1098 XH Amsterdam, The Netherlands}
\email{p.zhou@uva.nl}

\author{Xin Zhou}
\affil{Purple Mountain Observatory and Key Laboratory of Radio Astronomy, Chinese Academy of Sciences, 10 Yuanhua Road, Nanjing 210023, China}
\email{xinzhou@pmo.ac.cn}

\author{Yang Chen}
\affil{School of Astronomy and Space Science, Nanjing University,
163 Xianlin Avenue, Nanjing, 210023, China}
\affil{Key Laboratory of Modern Astronomy and Astrophysics, Nanjing University, Ministry of Education, PR China}

\author{Jie-Shuang Wang}
\affil{Tsung-Dao Lee Institute, Shanghai Jiao Tong University, Shanghai 200240, China}

\author{Jacco Vink}
\affil{Anton Pannekoek Institute for Astronomy, University of Amsterdam, Science Park 904, 1098 XH Amsterdam, The Netherlands}
\affil{GRAPPA, University of Amsterdam, Science Park 904, 1098 XH Amsterdam, The Netherlands}
\affil{SRON, Netherlands Institute for Space Research, Sorbonnelaan 2, 3584 CA Utrecht, The Netherlands}

\author{Yuan Wang}
\affil{Max Planck Institute for Astronomy, K\"{o}nigstuhl 17, 69117 Heidelberg, Germany}

\begin{abstract}
We have performed a multiwavelength study of
supernova remnant (SNR) \snr\ and its environment. The SNR hosts the magnetar \sgr, which emitted an extremely bright millisecond-duration radio burst on 2020 Apr 28 
\citep{chime20, bochenek20}.
We used the \twCO\ and \thCO~\Jotz\ data from the Milky Way Image Scroll Painting (MWISP) CO line survey to search for molecular gas associated with \snr, in order to constrain
the physical parameters (e.g., the distance) of the SNR and its magnetar.
We report that SNR \snr\ is likely impacting the molecular clouds (MCs) at the local standard of rest (LSR) velocity $\VLSR\sim 30~\km\ps$ and excites a weak 1720~MHz OH maser with a peak flux density of 47 mJy beam$^{-1}$. 
The chance coincidence of a random OH spot falling in the SNR is $\le 12\%$, and the OH--CO correspondence chance is $7\%$ at the maser spot. This combines to give $<1\%$ false probability of the OH maser, suggesting a real maser detection.
The LSR velocity of the MCs places the SNR and magnetar at a kinematic distance of $6.6\pm 0.7$~kpc.
The nondetection of thermal X-ray emission from the SNR and the relatively dense environment suggests \snr\ be an evolved SNR
with an age $t\gtrsim 1.6\E{4}(d/6.6~\kpc)~\yr$.
The explosion energy of \snr\ is lower than $2\E{51}(n_0/10~\cm^{-3})^{1.16}(d/6.6~\kpc)^{3.16}~\erg$, which is not very energetic even assuming a high ambient density $n_0=10~\cm^{-3}$. This reinforces the opinion that magnetars do not necessarily result from very energetic supernova explosions.

\end{abstract}


\section{Introduction} \label{sec:intro}

On 2020 April 28  at UTC 14:34:33 an extremely bright millisecond-duration radio burst was detected from the direction of \sgr\ by Canadian Hydrogen
Intensity Mapping Experiment CHIME in the 400--800 MHz band
\citep{chime20}.
Simultaneously, the Survey for Transient Astronomical Radio Emission
2 (STARE2) was triggered by this burst,  
and its fluence in the 1.4 GHz band was found to be $>1.5 $ MJy ms
\citep{bochenek20}.
The magnetar \sgr\ was previously detected in active states by the X-ray and $\gamma$-ray telescopes such as Swift, NICER, and Fermi-LAT
\citep{barthelmy20,fletcher20,palmer20,younes17}.
This radio burst was subsequently found to correspond to a hard X-ray burst, which was detected by INTEGRAL \citep{mereghetti20},
AGILE \citep{tavani20},
Insight-HXMT \citep{li20},
and Konus-Wind \citep{ridnaia20}. 
More specifically, Insight-HXMT detected a double-peaked hard X-ray counterpart from \sgr\ 8.57~s ahead of the radio double-peaked bursts. This unambiguously established a
relationship between the extraordinary fast radio burst (FRB)-like radio burst \citep{bochenek20,chime20}
and the magnetar, given that
there is no intrinsic delay between 
radio and X-ray bursts after a correction of the dispersion measure \citep[DM;][]{li20}.
While FRBs have been regarded millisecond-duration radio transients from cosmological distances \citep[see a recent review by][]{petroff19}, the new FRB or FRB-like burst from \sgr\ provides the first nearby, Galactic target for us to study in detail.

While many observational and theoretical studies have been underway 
for this particular burst event in our Galaxy, 
the current understanding of the host of \sgr\ --- supernova remnant (SNR) \snr\ \citep{sieber84} --- is still limited.
The distance, age, and explosion properties are shared between
the magnetar and SNR. Therefore, the study of \snr\ provides essential information for the magnetar and also its radio bursts.
The association between \snr\ and \sgr, located in its geometric center, was only proposed recently \citep{gaensler14}, shortly after the discovery of \sgr\ 
\citep{cummings14}.
The small characteristic age of \sgr\ \citep[3600~yr;][]{isarel16} also supports that its SNR should be visible.

There have been many disputes on the distance of \sgr. 
Most of the distance measurements were 
targeted to SNR \snr, but from the blackbody emission of SGR 1935+2154, \cite{kozlova16} estimated an upper limit of the distance to be $<10.0$ kpc.
An assumed distance of 9~kpc was adopted for
the magnetar \sgr\ by \cite{isarel16} and \cite{younes17}, who referred to the presumed distance of \snr\ from 
the empirical relation of radio surface-brightness -- distance ($\Sigma$--$D$) for SNRs \citep{pavlovic14},
a method with a large intrinsic scatter.
The distances of $\sim 7$~kpc and $14.3$~kpc were proposed in
different studies using this relation \citep{park13a,pavlovic14}.
The lower limit of the distance (4.5~kpc) has been constrained 
from  the HI absorption feature toward the SNR
at the local standard of rest (LSR) velocity $\VLSR\sim 40~\km\ps$
\citep{kothes18,ranasinghe18}.
Overall various estimates are obtained  based on the HI-absorption
method, from a far distance of $11.7\pm 2.8~\kpc$  \citep{surnis16} to a closer distance of 4.5--9~kpc \citep{ranasinghe18}.
\cite{kothes18} favored a distance of $12.5\pm 1.5$ kpc, due to a morphological match between an HI cavity at $\VLSR=-51$ to $-44~\km\ps$ and the SNR. 
The distance inferred from the dispersion measure is $\sim 9$ or 15~kpc, varying with electron distribution models \citep{kothes18,zhong20}.

Motivated by the uncertain distance and poorly known SNR properties of \snr, and their potential utilization in the understanding of the radio burst occurring on 2020 Apr 28, we performed a molecular environment study of \snr.
We show here that the SNR is likely associated with a molecular cloud (MC), which helps to constrain the distance 
by comparing 
the LSR velocity 
with the Galactic rotation curve \citep[e.g.,][]{reid14}.
We also revisited the multiwavelength data to constrain the SNR properties such as the SNR age and explosion energy, which are the shared properties for the magnetar \sgr.
Simultaneously with this paper, \cite{mereghetti20} reported an independent distance measurement (3.1--7.2~kpc) using the dust scattering X-ray halo around \sgr, which covers our suggested value of $6.6\pm 0.7$~kpc.

\section{Data in the Multiwave bands}

We obtained \twCO\ \Jotz\ and \thCO~\Jotz\ data from the Milky Way Image Scroll Painting (MWISP\footnote{http://english.dlh.pmo.cas.cn/ic/})--CO line survey project.
The project is an unbiased high-sensitivity survey toward the Galactic plane using the Purple Mountain Observatory (PMO) Delingha 13.7~m millimeter-wavelength telescope 
with a $3\times3$ multibeam sideband separation superconducting receiver \citep{shan12} as the front end and a fast Fourier transform spectrometer (1~GHz total bandwidth) as the back end.
The half-power beamwidth was about $55''$.
The typical rms noise level is $\sim$0.5~K for \twCO~(\Jotz) in a 0.17~\kms\ channel and $\sim$0.3~K for \thCO(\Jotz) in a 0.16~\kms\ channel.
A detailed description of the observation can be found in \cite{zhou16c}.
All data were reduced using the GILDAS/CLASS package.

We retrieved the Very Large Array (VLA) radio continuum image at 1.4~GHz, the data cubes of 1720, 1667, 1665 and 1612~MHz OH lines, and HI data from the HI/OH/recombination line survey (THOR) project \citep{beuther16, wang20}.
The combined THOR plus the VLA Galactic Plane Survey (VGPS) 1.4~GHz image provides a spatial resolution of $25''$.
The OH data cubes have an angular resolution of $\sim 12''$ and a velocity resolution of $1.5~\km\ps$.
The LSR velocity range of data cube spans from $-58.5~\km\ps$ to 135~\kms.
We also retrieved the \Spitzer\ $24 \um$ post-basic-calibrated data from the \Spitzer\ archive. The mid-IR observation was performed as a 24 Micron Survey of the Inner Galactic Disk Program (PID: 20597; PI: S. Carey).

\begin{figure*}
\epsscale{1.15}
\plotone{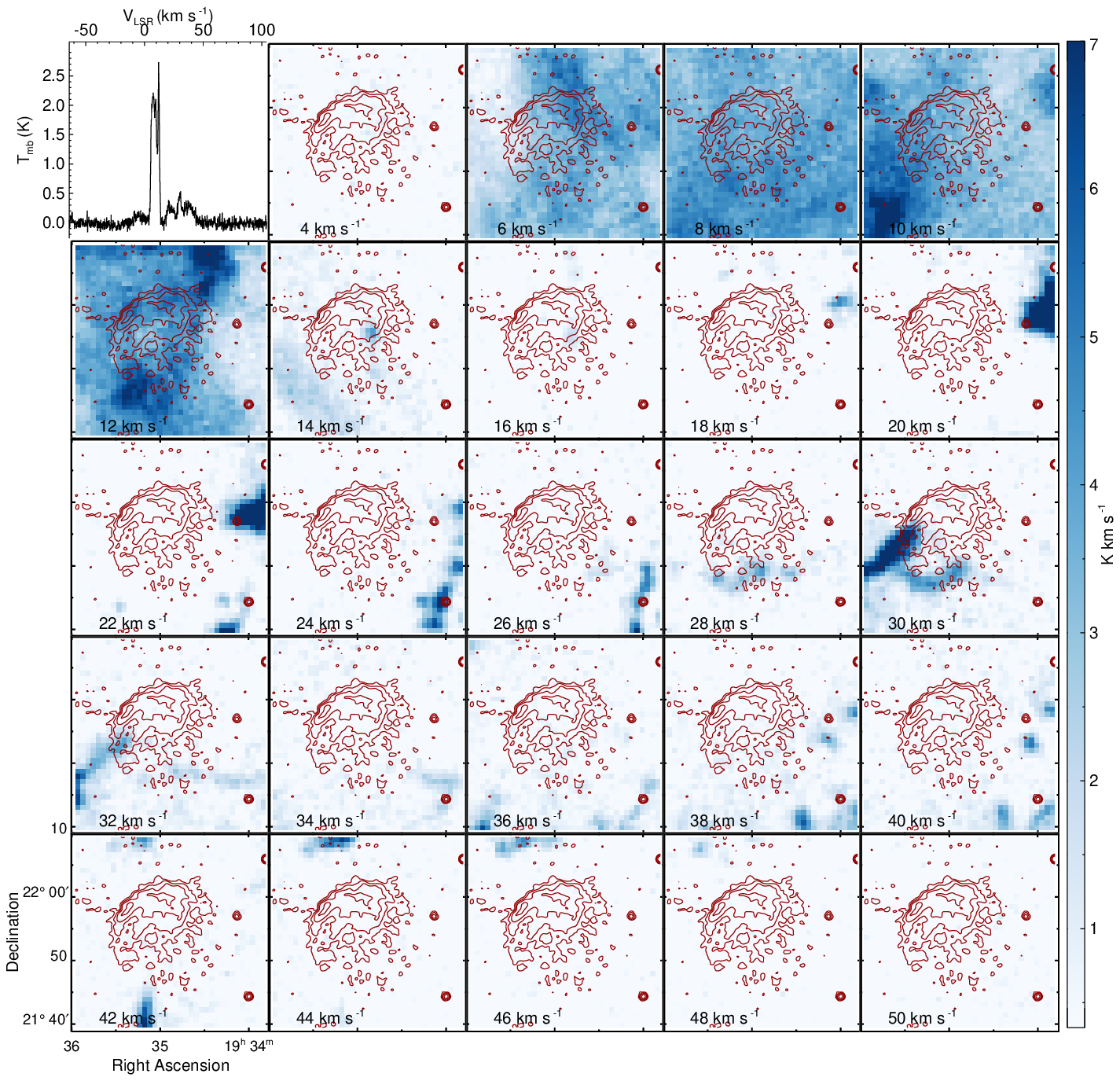}
\caption{
A grid of the velocity-integrated intensity maps of PMO \twCO~\Jotz\ emission with a velocity step of 2~\kms.
The contours are taken from the THOR 1.4~GHz radio continuum.
The first panel shows the \twCO~\Jotz\ spectrum averaged across
the field-of-view.
\label{fig:cogrid}}
\end{figure*}

We revisited all the available \XMMN\ data 
of \snr\ 
to search for its extended X-ray emission. 
The archival \Chandra\ observations were not used as they covered only a small fraction of the SNR.
\snr\ was observed with \XMMN\ in five epochs during 
2014 and 2015 (OBSIDs: 0722412801, 0722413001, 0748390801,
0764820101, and 0764820201, PIs: G. Israel or N. Schartel).
After removing the high background periods from the
events, the screened exposure time of pn, MOS1 and MOS2 are
53~ks, 93~ks, 85~ks, respectively.
The \XMMN\ data were reduced using the Science Analysis 
System software 
(SAS, vers.\ 16.7.0).
Finally, we obtained  the background-subtracted, vignetting-corrected, and adaptively smoothed image of \snr, with all of the pn and MOS images combined.

\section{Results}

\subsection{Molecular environment of \snr} \label{sec:co}

The overall \twCO~(\Jotz) spectrum in the field-of-view (FOV) shows several velocity components, 
$\VLSR=-10$--15~\kms\ and $\VLSR=15$--50~\kms\ (see the first panel in Figure~\ref{fig:cogrid}). 
There is also faint CO emission at around $\VLSR=-40$~\kms, 
but it is too weak to be discerned in the overall spectrum and the emission is found 
outside the SNR boundary (see Figure~\ref{fig:cogrid_negv} in the Appendix).
Figure~\ref{fig:cogrid} shows the distribution of \twCO\ emission toward \snr\ from $\VLSR=4~\km\ps$ to $50~\km\ps$.
The morphological overlap between the SNR and MCs has been only found in two velocity ranges, $\VLSR=6$--$14~\km\ps$ and 30~\kms.

The strong and diffuse CO emission at $\VLSR=6$--14~\kms\ 
is probably from nearby MCs at $\sim 1$~kpc. 
However, this velocity also corresponds to a far distance of $\sim 8$~kpc according to the Galactic rotation curve \citep{reid14}. 
At $\sim 12~\km\ps$, the \twCO\ intensity map shows a weak gradient from
the east to the west and some enhancement along the remnant eastern edge.
The gradient and enhancement might explain the morphology of radio emission of \snr\ in principle.
However, 
no physical evidence \citep[e.g., line broadening, heating, 1720 MHz OH masers;][]{jiang10, chen14} of SNR--MC interaction is found in the velocity range of $6$--14~\kms,
in which multiple velocity components are crowding.

\begin{figure}
\epsscale{1.1}
\plotone{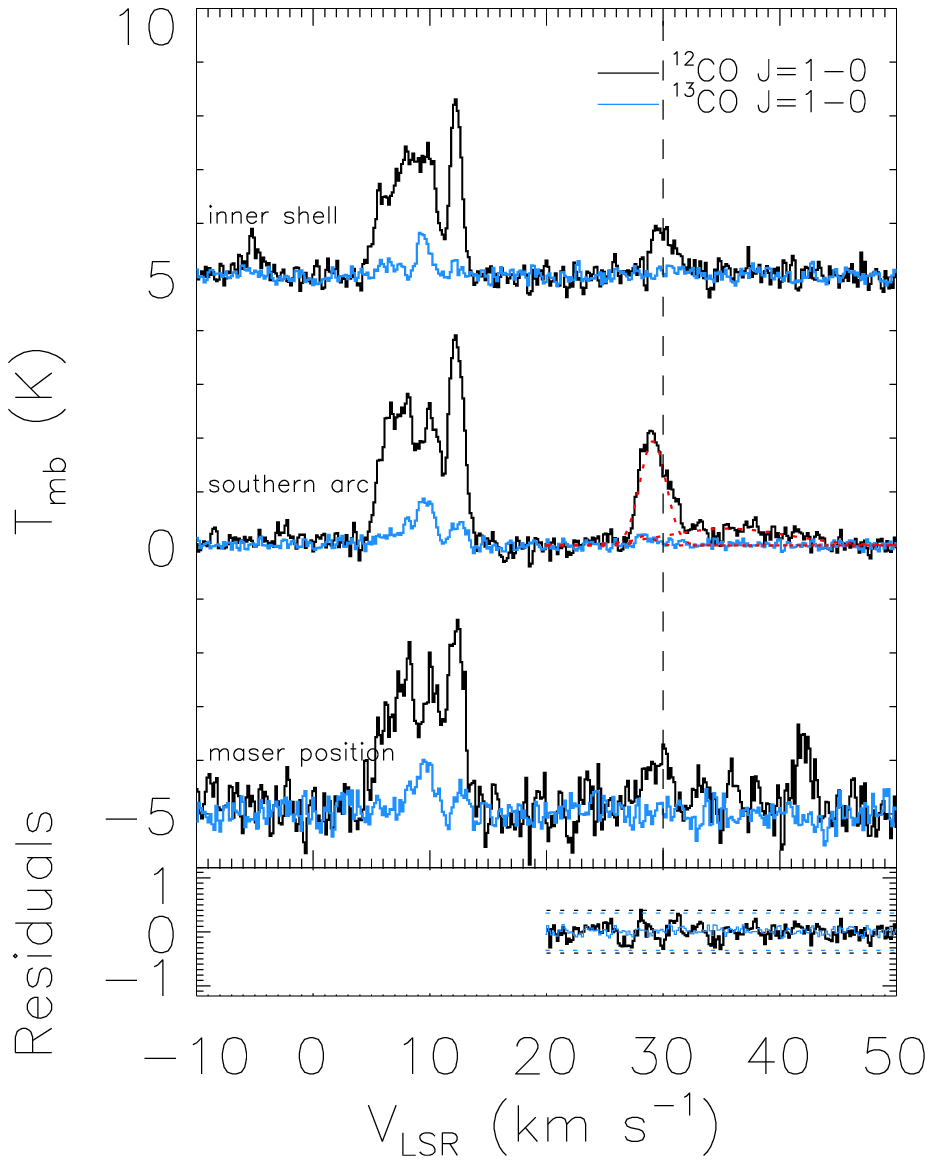}
\caption{
Spectra of the \Jotz\ transitions of \twCO\ and \thCO\ 
from the inner shell, southern arc-like region, and the maser position (see the regions labeled in Figure~\ref{fig:radio_co_24um}).
The vertical dashed line denotes $\VLSR=30~\km\ps$.
The residual panel is given for the southern arc spectra fitted using two Gaussian lines (red dashed lines).}
\label{fig:spec_co}
\end{figure}

\begin{figure}
\plotone{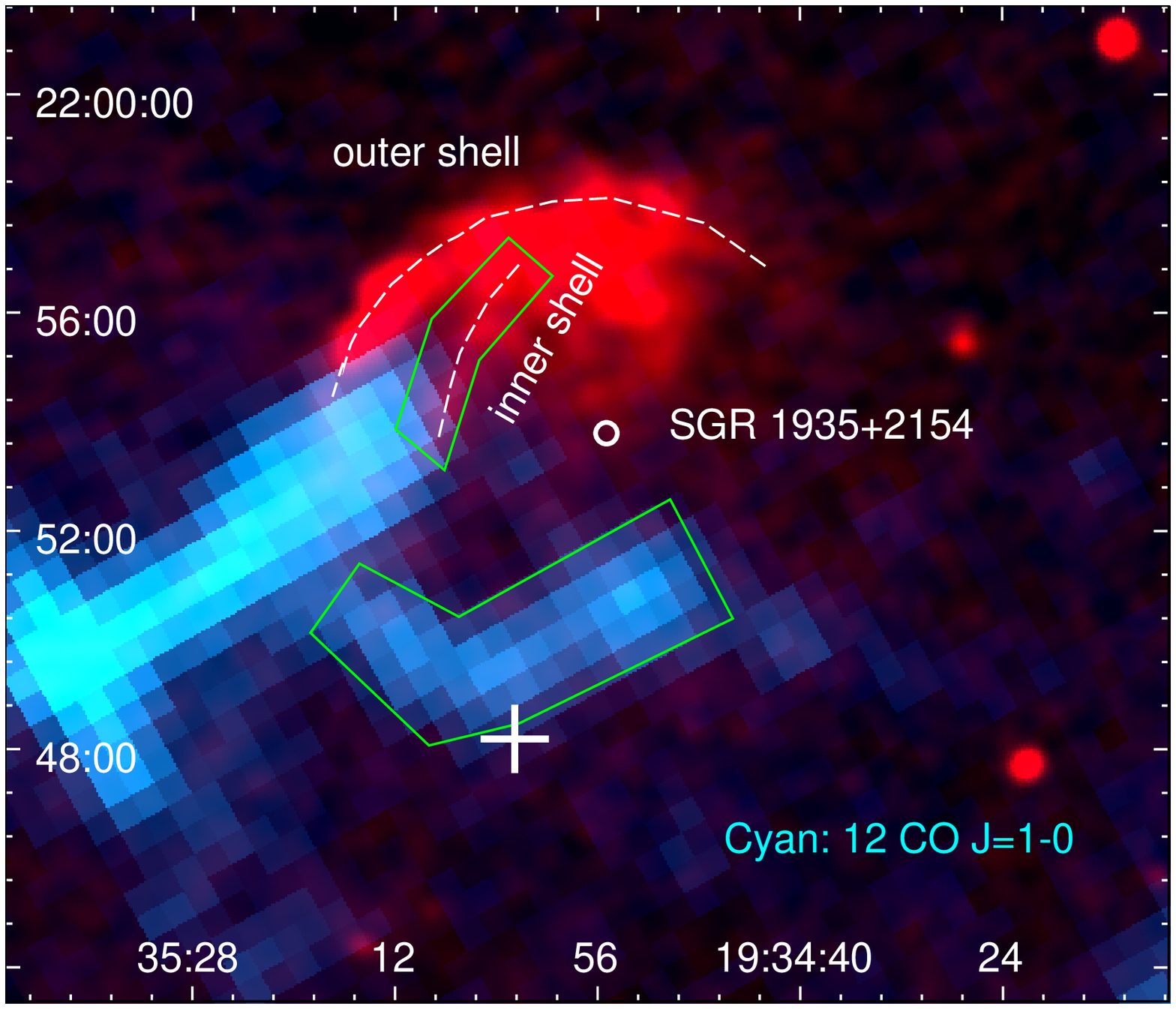}
\plotone{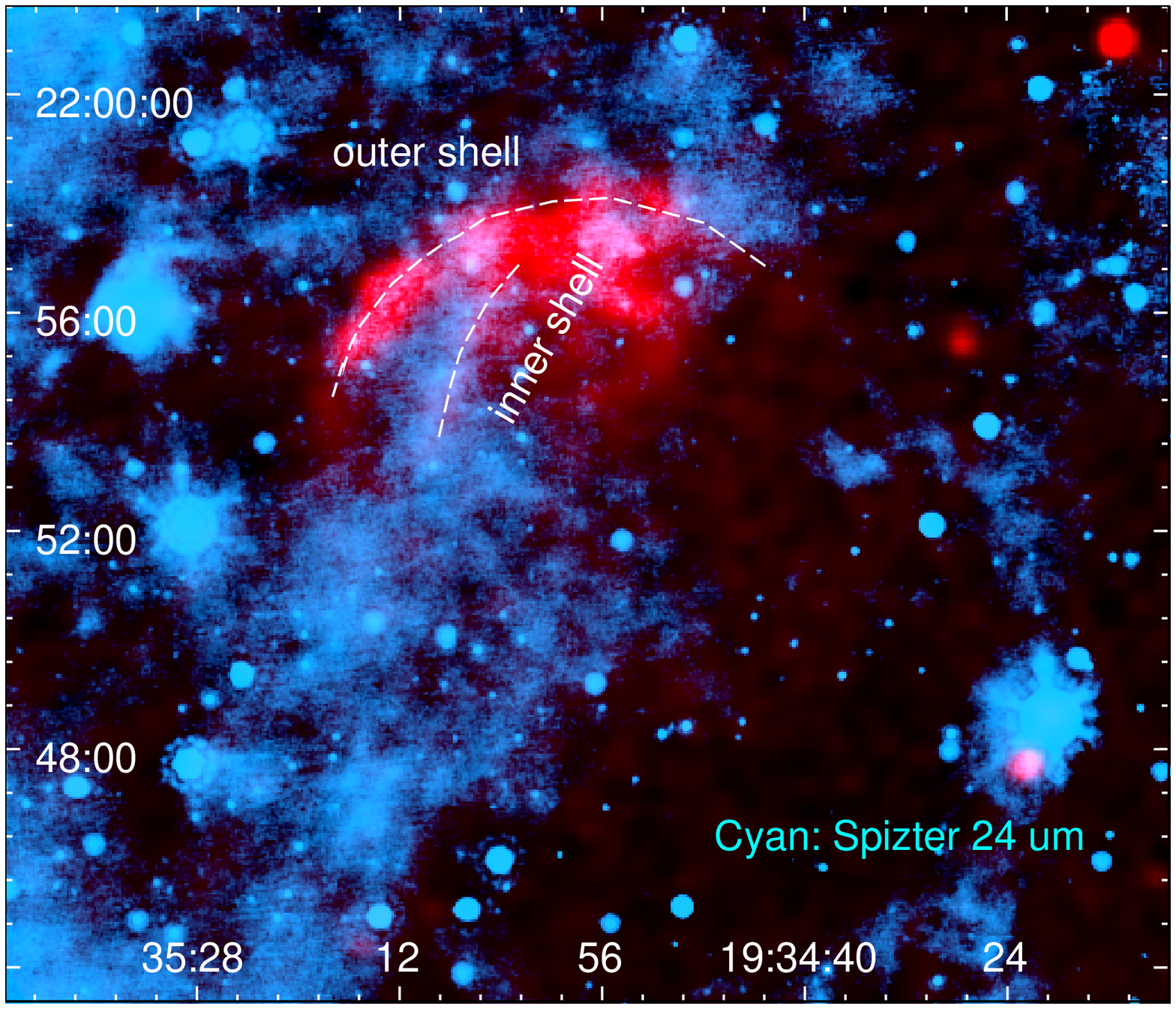}
\caption{
Composite images comparing the THOR 1.4~GHz radio continuum (red) and
PMO \twCO~\Jotz\ (upper panel, cyan color; $\VLSR=28$--32~\kms) or \Spitzer\ $24~\um$ (lower panel, cyan color) emission.
The polygons in the upper panel define the regions ``inner shell'' and ``arc'' for CO spectral extraction.
The circle and cross signs denote the position of the magnetar
and 1720~MHz OH maser.
\label{fig:radio_co_24um}}
\end{figure}

At $\VLSR\sim 30~\km\ps$,  there are two prominent molecular features
near \snr,
a molecular bar connecting to the radio shell and 
an arc-like structure in the SNR south.
The \twCO\ and \thCO\ spectra from the arc-like structure are presented in Figure~\ref{fig:spec_co} and the extraction region is shown in the upper panel of Figure~\ref{fig:radio_co_24um}.
The \twCO\ line profile contains a narrow component at $\VLSR\sim 30~\km\ps$ and a broad wing in the range of 30--45~\kms,
while a weak \thCO\ emission is only seen at the narrow component.
We fitted the two \twCO\ components with two Gaussian lines, 
providing full width half maximum (FWHM) line widths of 2.6$\pm0.1$~\kms\ and 12.1$\pm1.3$~\kms for the narrow and broad lines, respectively.
For the broad \twCO\ component, the line centroid and intensity are
$35.5\pm 0.6~\km\ps$ and $0.31\pm 0.02~\K$, respectively.
For the narrow \twCO\ component, the line centroid and intensity are
$29.12\pm 0.03~\km\ps$ and $1.94\pm 0.06~\K$, respectively.
Using more Gaussian components does not significantly improve the fitting, as
there is no extra component notable in the residuals.

The existence of a broad \twCO\ component with a width of $dV\sim 12~\km\ps$ indicates an interaction between the remnant and the MC,
as a line broader than typical molecular lines  requires external perturbation.
According to Larson's third law \citep{larson81}, the MC velocity dispersion $\sigma$ is correlated to the cloud size $L$, $\sigma (\km\ps)= 1.1 L (\pc)^{0.38}$. This suggests that the typical
molecular line width $dV=2.355\sigma$ is within a few $\km\ps$.
Nevertheless, given the low sensitivity of the \thCO\ emission, we do not know whether the \twCO\ emission is indeed optically thin (e.g., with a large \twCO/\thCO\ line ratio) or not.
We saw line crowding in some regions in the FOV, where multiple
line components are shown between $\VLSR=20$ and $45~\km\ps$.
For this reason, although the broad line at the molecular arc
appears to be a single Gaussian, future molecular observations 
will be needed to test the line-crowding possibility. 
Therefore, based on this alone, we cannot rule out a broadening due to line-of-sight effects.
We also plot in Figure~\ref{fig:spec_co} the \twCO\ and \thCO\ spectra from the inner shell region defined in Figure~\ref{fig:radio_co_24um}. We have not found any broadened
line in this region or other parts of the FOV. 
The narrow line in the inner shell and the maser point is about 2 times weaker 
than that from the arc-like structure.  With the fainter CO emission, the potential 
broadened line emission is under our detection limit.

The CO gas at $30~\km\ps$ corresponds to either a near distance of $d\sim 2.5$~kpc or a far distance of $\sim 6.6$~kpc.
Since the SNR has been established to be farther than 4.5 kpc, an association between the CO and the SNR implies a distance of $\sim 6.6$ kpc.
Using the assumption of local thermal equilibrium, we estimated the H$_2$ column density of the $\VLSR\sim 30~\km\ps$ narrow component to be $N({\rm H_2})\simeq2.8\E{20}$~cm$^{-2}$ and the mass to be $M({\rm H_2})\simeq200\du^{2}$~$\Msun$, where $\du$=$d$/(6.6~kpc) \citep[see ][]{zhou16c}.
Assuming that the size of the \twCO\ arc in the line-of-sight is the same as its width, the density of the narrow component is $n$(H$_2$)=$N$(H$_2$)/depth$\simeq20\du^{-1}~\cm^{-3}$.
The low mean density suggests that the MCs are highly clumpy and not resolved in the CO
observations.

In Figure~\ref{fig:radio_co_24um}, we compare the molecular structures with the radio image of \snr\ and the mid-IR emission.
In the radio band, \snr\ is limb brightened and has a faint spherical 
halo in the south \citep{sieber84, kothes18}.
There is more than one explanation for the limb-brightened radio morphology.
It could result from 
a density enhancement of the ambient medium in the northeast, which
is, however, not seen in our molecular maps or HI map at $\VLSR\sim 30~\km\ps$ (except for the inner shell, see the HI image in Figure~\ref{fig:HI} and relevant discussion in Section~\ref{sec:HI} in the Appendix C).
Although the CO emission at $\VLSR\sim 12 \km\ps$ is enhanced in the radio-brightened hemisphere of the SNR, we have not found physical evidence to support its association with the SNR. \cite{kothes18} found an HI
cavity at $\VLSR\sim -46~\km\ps$ morphologically enclosing the northeastern hemisphere of the SNR, also in need of kinematic evidence.
The radio morphology could also result from an enhancement of the magnetic fields on one side \citep{orlando07,west16}
or interaction with the winds of a runaway progenitor star \citep{zhang18}.

The spherical radio morphology indicates that
SNR recently impacts with dense medium, which could 
be dense enough to deform the SNR morphology.
The southern molecular arc does not confine the radio halo, 
but could be a structure behind the SNR, as indicated by the possible redshifted broad line (see Figure~\ref{fig:spec_co}).
The redshifted velocity is $\sim 6~\km\ps$, lower 
than the blast wave velocity $\vs\lesssim 200~\km\ps$ (see Section~\ref{sec:snr}) and the cloud shock velocity $v_{\rm c}\sim 12~\km\ps$ implied by the broad \twCO\ line.
Assuming a ram pressure balance between the blast wave and the cloud shock, we have $n_{\rm c} v_{\rm c}^2 \sim n_0 v_s^2$ \citep{mckee75}, where $n_{\rm c}$ and $n_0$ are the shocked cloud density and preshock intercloud density, respectively.
Therefore, the cloud shock velocity is correlated to the density ratio between the preshock intercloud and shocked dense cloud: $v_{\rm c}\lesssim 200 (n_0/n_{\rm c})^{1/2} \km\ps$.
Although the density ratio is unclear ($< 1$), both $v_{\rm c}$ and redshifted velocity values are in reasonable ranges.

The THOR radio image shows a relatively faint inner radio shell in the north, 
well correlated with the mid-IR $24~\um$ emission (see the lower panel of Figure~\ref{fig:radio_co_24um}).
The spatial correlation between the two bands has been found 
in a number of SNRs that interact with dense medium \citep[see][and references therein]{pinheirogoncalves11,koo16}.
The existence of an inner shell structure 
also favors an impact of shock with the dense gas in the foreground or background, similar to those found in SNR W28 \citep[e.g.,][]{frail94, claussen97, arikawa99,zhou14}.
As shown in Figure~\ref{fig:radio_co_24um}, the eastern molecular bar ends at the inner radio shell, consistent with the picture that 
an interaction with dense medium causes the inner shell.

\subsection{1720~MHz OH maser}

\begin{figure}
\epsscale{1.1}
\plotone{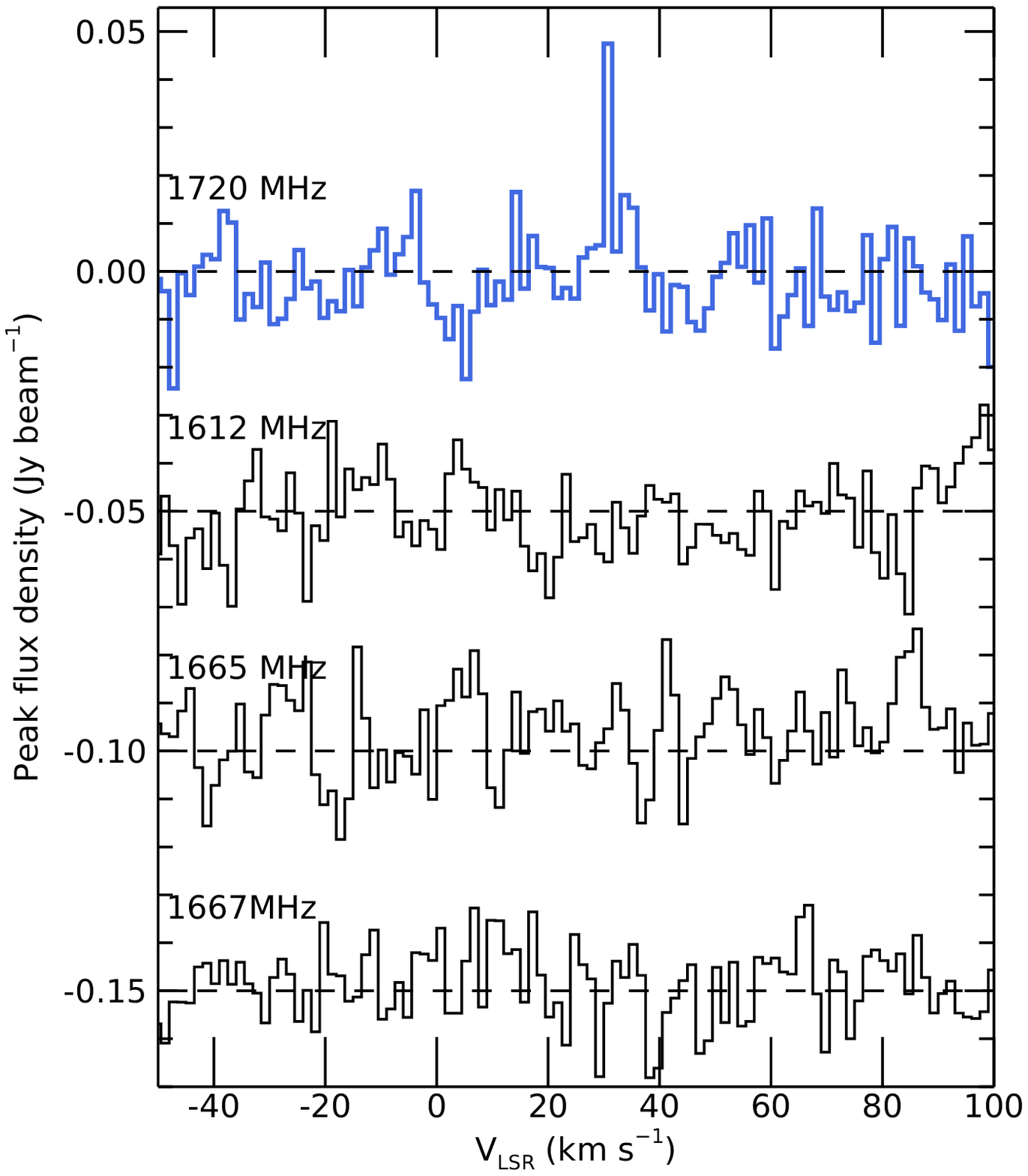}
\caption{
Spectra of four ground-state transitions of OH at
the position $\RAdot{19}{35}{02}{84}$, $\decldot{21}{48}{11}{59}$ (J2000; see the cross
in Figures~\ref{fig:radio_co_24um} and \ref{fig:radio_x}).
\label{fig:oh}}
\end{figure}

1720~MHz OH masers, when unaccompanied by the other ground-state transitions at 1662, 1665, and 1667 MHz that are usually pumped by a far-IR radiation field, are regarded as signposts of the
collision between shock and MCs
\citep[e.g.,][]{frail94,claussen97,wardle02}.
Compared to the thermally excited OH emission, OH masers
are narrow \citep[generally $\lesssim 2~\km\ps$;][]{claussen97}, compact, and too bright to be explained as thermal lines.

From the THOR OH survey, we have found a 1720~MHz line
at the southern radio boundary of \snr\ at $\VLSR=30~\km\ps$ (see Figure~\ref{fig:oh},  from the OH tile centered at $l=56.^\circ 75$).
The line is identified from a spatially unresolved region
and shown in two tiles of OH data 
cleaned separately.
The cleaning areas of the two tiles 
center at $l=56.^\circ 75$ and $58^\circ$, respectively, 
with a size of $4^\circ\times 2.5^\circ$, while 
other cleaning parameters are the same and the data were from the same survey \citep[see][for details]{beuther16}.
The line is also narrow, as the line profile is unresolved with the velocity resolution of $1.5~\km\ps$. 
Moreover, we have not found accompanied 1612, 1665, and 1667~MHz OH lines at the position and LSR velocity.
The peak flux densities of the 1720~MHz OH line are 47 and 44~mJy beam$^{-1}$ in the tiles centered at $l=56.^\circ 75$ and $58^\circ$, respectively, corresponding to a high radiation temperature of $\sim 130$~K
for a given beamwidth of $12''$.
The radiation temperature is significantly larger than the \twCO\ temperature along the line-of-sight.
All of these properties suggest that the line is a collisionally excited OH maser.

The OH maser was not reported before, probably because previous surveys did not have the required sensitivity \citep[25 mJy in][]{hewitt09} or the line is too narrow.
We note that 1720~MHz maser is only detected in a single velocity-channel, which could be caused by
the narrow width of the OH maser. 
The maser was not listed in the OH maser catalog compiled in \cite{beuther19}, which selected masers with $>2$-rms detection in more than one velocity-channel. 
The two-channel criterion in \cite{beuther19} is used to 
reduce sporadic false detection and ensure clean detection.
However, it could result in incomplete detection because it tends to omit narrow and faint masers.
Narrow 1720~MHz OH masers with width $\le1.5~\km\ps$ have been extensively
found in SNRs.
\cite{claussen97} show that around half of the OH masers 
in W44 and W28 are narrower than $1~\km\ps$ or detected 
in less than three channels (channel resolution of 0.53~\kms).
The 1720~MHz OH line near \snr\ 
matches all the other criteria in \cite{beuther19}, given its $\gtrsim 5$-rms intensity at a single position and 
its brightness of $>20$~mJy beam$^{-1}$ in over 12 pixels (pixel 
size of $3''$). 
The rms value is determined using the OH spectrum at the maser spot  (see Appendix~A for details.)
Moreover, the maser's position is coincident with the radio boundary of \snr\ and its velocity is consistent with that of the \twCO\ arc. 

In Appendix~A , we provided details of the maser identification and its likelihood.
We have calculated the chance coincidence of an OH 
spot randomly falling in the SNR, using the real 1720, 1612, 1665, and 1667 MHz data 
in a sky region 16.4 times that of  the SNR area. We considered
the rms distribution in the spatial and channel
dimensions. 
Subsequently, we calculated how frequently
an OH spot at a random velocity corresponds
to \twCO\ emission.
A false OH spot would be found at any velocity channel and does not need to correspond to \twCO\  emission.
The chance coincidence of a maser spot in the 
SNR is $\le 12\%$ at the detected maser's significance level, and the chance of OH--CO 
correspondence is 7\%. 
The two probabilities are multiplied to give the probability for a 
false OH spot in the SNR that happens to correspond to MCs.
As a result, the false detection probability
is $< 1\%$,
suggesting a real detection. 

The maser position is located at the outer boundary of the molecular arc and lies away from the radio-bright shell (see Figure~\ref{fig:radio_co_24um}a).
The location could be due to the projection effect or/and a favorable excitation condition \citep{lockett99}.
Masers are highly beamed emission. 
It is more likely to observe masers near the SNR boundary, because
the collisionally excited masers are beamed toward the observer who views the shock ``edge-on'' \citep[e.g.,][]{hollenbach13}.
The angular resolution of the PMO \twCO\ data is around $1'$ ($\sim 1\du$~pc), much larger than 
the maser or maser groups \citep[10--$10^3$~au,][]{elitzur92,beuther19}. 
The \twCO\ emission across the SNR is much weaker than the typical dense MCs \citep[$\sim 10$~K; see, e.g.,][]{drain11}, indicating that the gas is highly clumpy and unresolved using the PMO observation. 
Consequently, the beam dilution causes a weak CO emission.
This also prevents us from deriving the temperature, density, and column density of the molecular core that produces the OH maser. 
Figure~\ref{fig:spec_co} shows that our PMO CO observation is not sensitive enough to detect weak, broadened CO emission at the maser point.
Future high-resolution and high-sensitivity observations are needed to search for
broadened molecular emission at the maser point.
Nevertheless, earlier 1720 MHz OH maser surveys in SNRs have revealed that the masers 
do not always correspond to the radio or CO peaks. Instead, they could appear at radio-faint structures \citep{frail96,hewitt09} or CO-weak regions \citep[e.g., 3C391, N49;][]{frail96,brogan04}.
A cluster of masers has also been found in
the outer boundaries of the MCs in W28 and W44 \citep[see Figure 1 in][]{frail98}. 
The CO and maser emission together provide evidence for an SNR--MC interaction, with the SNR being located at $\sim 6.6$ kpc.

\subsection{X-ray analysis }

\begin{figure}
\epsscale{1.1}
\plotone{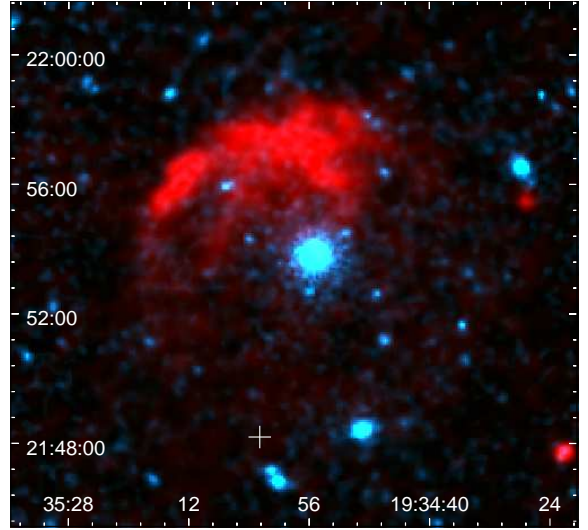}
\caption{
A composite image of \snr\ in the \XMMN\ X-ray (cyan; 0.4--7.2 keV) and
THOR 1.4~GHz radio bands (red). 
\sgr\ is the bright X-ray source in the center.
The cross sign in the south denotes the position of the 1720~MHz OH maser.
\label{fig:radio_x}}
\end{figure}

Figure~\ref{fig:radio_x} shows the 0.4--7.2 keV X-ray emission of \snr\ and \sgr, with the VLA 1.4~GHz continuum coded in red for comparison.
We have not found any evidence of the extended X-ray emission 
from the shell or the interior of the SNR. 
Considering that the SNR is interacting with a dense medium, the nondetection of X-ray emission is not because of the low gas density, but suggests that the shock velocity
is too low to heat the gas to the high temperature.

\section{Discussion}
\subsection{Distance}

We have shown that \snr\ is likely associated with CO gas and excites a 1720~MHz OH maser at $\VLSR\sim 30~\km\ps$.
This LSR velocity corresponds to a kinematic distance of $6.6\pm 0.7$~kpc according to the rotation curve of the Galaxy \citep{reid14}, where the uncertainty is given  at the 1 $\sigma$ level based on the Monte Carlo method by \cite{wenger18}.
The distance falls within the lower and upper limits (4.5--10~kpc) constrained using the HI-absorption method \citep{ranasinghe18} and the magnetar's blackbody emission \citep{kozlova16}, respectively.

\subsection{SNR properties} \label{sec:snr}

We herein update the SNR age $t$ using the new distance.
The nondetection of the X-ray emission (see Figure~\ref{fig:radio_x}) supports that the SNR gas has significantly cooled.
Therefore, it is reasonable to consider the SNR has reached the radiative phase,
as suggested by \cite{kothes18}.
During this phase, the SNR expansion is pressure driven, having a solution 
$R_{\rm s}\propto t^{0.31}$ \citep{chevalier74}, where
the SNR radius $\Rs=10.6\du~\pc$ for
an angular radius of $5\farcm{5}$. 
Using an upper limit of the shock velocity $\vs$ of $200~\km\ps$ in 
the radiative phase \citep{vink12}, the SNR
age is estimated as $t=0.31 \Rs/\vs> 1.6\E{4}\du~\yr$.
This is still larger than the characteristic age of \sgr\ obtained from the spin properties \citep[3600~yr;][]{isarel16}.
However, as pointed out in previous studies \citep[e.g.,][]{olausen14, kothes18}, the characteristic ages are poor age indicators, while SNR ages better represent the real ages of the magnetars.

The explosion energies of magnetars in SNRs have been found to vary by over one order of magnitude \citep{vink06c,zhou19}. Particularly, SNR RCW~103 hosting the magnetar 1E~161348$-$5055 originated from a very low energy explosion \citep[$\sim 10^{50}~\erg;$][]{zhou19,braun19}.
Therefore, we cannot simply assume a canonical explosion energy $E=10^{51}~\erg$ for \snr\ and \sgr. 
The explosion energy of an SNR in the radiative phase can be estimated as 
$E=1.5\E{50}n_0^{1.16} (\Rs/10.6~\pc)^{3.16} (\vs/200~\km\ps)^{1.35} \erg$ \citep{cioffi88}, 
%$E=1.4\E{50}n_0^{1.12} (\Rs/10.7~\pc)^{3.12} (\vs/200~\km\ps)^{1.4} \erg$ \citep{chevalier74}, 
where $n_0$ is the mean density of the ambient gas.
We do not expect the $n_0$ value to be as large as the MC density $\nHH\sim  20~\cm^{-3}$.
The progenitors of magnetars are massive stars, whose stellar winds can create large low-density bubbles before the supernova explosions \citep{chevalier99, chen13}.
As shown in Figure~\ref{fig:radio_x},
the radio morphology of \snr\ is almost round (although limb brightened) and MCs overlap only a small portion of the SNR.
This is consistent with the scenario that the SNR was initially evolved in a homogeneous, low-density medium until it reached the dense molecular gas.
If we use a large mean ambient gas density $n_0=10~\cm^{-3}$, 
we derived the explosion energy of \snr\ as $<2.1\E{51} \du^{3.16} \erg$.
Given the distance and low SNR velocity, \sgr\ is unlikely to be formed from a very energetic explosion, consistent with those in other magnetars in SNRs \citep{vink06c,martin14,zhou19}.

\section{Conclusion and remarks}
Here we summarize the results from the multiwavelength study and
our concerns about the uncertainties of the observational results.
We have performed molecular environment study of \snr\ and found that the MCs at
$\VLSR=6$--14~\kms\ and 30~\kms\ are spatially overlapping the SNR.
The physical interaction between the SNR and MCs is mainly built on a single, weak 
1720~MHz maser detected at $\VLSR=30~\km\ps$, as 1720~MHz OH masers are regarded as signposts of
shock--cloud interaction.
Moreover, the spatial correlation between the inner shell and mid-IR emission, and the existence
of MC connecting to the inner shell at $\VLSR\sim 30~\km\ps$ provides additional morphological support.
Nevertheless, we are aware that further high-resolution molecular observation is needed to provide more kinematic evidence and confirm this association.
We have analyzed the \XMMN\ X-ray data to search for thermal X-ray emission from
\snr.
The nondetection of X-ray emission suggests that the SNR is old.

Based on the LSR velocity of the MCs, we derived a kinematic distance  of \snr\  as $ 6.6\pm 0.7~\kpc$.
We also constrained the SNR age to be larger than $1.6\E{4}\du^{-1}~\yr$ and the supernova explosion energy to be less than $2.1\E{51}\du^{3.16} \erg$.
These properties are shared between \snr\ and \sgr.

\begin{acknowledgements}

We thank Hao Qiu for discussion on FRBs and thank
Sera Markoff and Vladim{\'i}r Dom{\v c}ek for helpful comments.
P.Z.\ acknowledges the support from the Nederlandse Onderzoekschool Voor Astronomie (NOVA) and NWO Veni Fellowship.
This work is supported by National Key R\&D Program of China grants Nos.\ 2017YFA0402701  and 2017YFA0402600, and 
Key Research Program of Frontier Science, CAS, grant No.\ QYZDJ-SSW-SLH047, and NSFC grants Nos., 11503008, 11590781, 11403104,
11773014, 11633007 \& 11851305.
J.S.W.\ is supported by China Postdoctoral Science Foundation. 
Y.W.\ acknowledges support from the European Research Council under the Horizon 2020 Framework Program via the ERC Consolidator Grant CSF-648505.
\end{acknowledgements}

\software{
GILDAS,\footnote{https://www.iram.fr/IRAMFR/GILDAS/}
SAS,\footnote{https://www.cosmos.esa.int/web/xmm-newton/download-and-install-sas}
DS9\footnote{http://ds9.si.edu/site/Home.html},
Starlink \citep{currie14}
}

\appendix

\section{The reliability of the 1720~MHz OH maser detection}  \label{sec:iden}

The 1720~MHz OH maser at 
the position ($\RAdot{19}{35}{02}{84}$, $\decldot{21}{48}{11}{59}$, J2000) is narrow and weak, with a peak flux density of 47~mJy beam$^{-1}$ in the data tile centered at $l=56.^\circ 75$. It is found at the boundary
of the SNR and coincides with CO emission.
The cleaning setup influences the noise distribution of the interferometric data.
The cleaning-introduced noise could disappear in other data with a different cleaning area.
We took the neighboring tile at $l=58^\circ$ 
and also detected the maser at the same position and velocity with a peak flux density of 44 mJy beam $^{-1}$.
The data of the two tiles were taken from the same survey, but the areas used for cleaning shifted by $1.^\circ 25$.
As the tiles at $l=58^\circ$ and $l=56.^\circ{75}$ (the one used in the paper) were cleaned separately, it is highly unlikely the cleaning-introduced noise spot would be at the same position and velocity.
Hereafter we calculate the probability for a false detection randomly coincident with the SNR.

\begin{figure*}[h]
  \centering
\includegraphics[angle=270, width=0.8\textwidth]{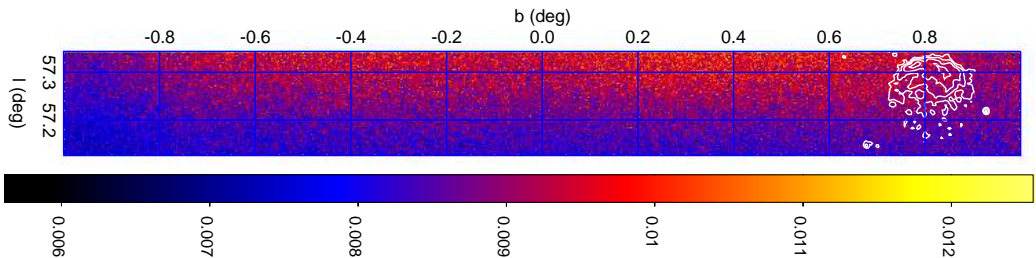}
\caption{
The rms distribution of the 1720~MHz OH data in the large-scale region centered at ($l=57.^\circ 235$, $b=0^\circ$) with a size of $13'\times 2^\circ$.
The radio contours of \snr\ are overlaid.
The unit of the color bar is Jy beam$^{-1}$.
The Galactic longitude ($l$) and latitude ($b$) axes are rotated.
}
\label{fig:rms_1720}
\end{figure*}

We first identified OH spots from a large-scale region centered at ($l=57.^\circ 235$, $b=0^\circ$) with a size of $13'\times 2^\circ$. 
We selected a region along the latitude because the rms along the longitude has a large variation (see Figure~\ref{fig:rms_1720}), and because identifying clumps in a very large region requires computing resources.
Two tiles with different cleaning areas are cross-matched to filter out 
cleaning-introduced noise.
The identified clumps are used to give the probability for an OH clump falling in the SNR ($P_1$).
Subsequently, we examined the \twCO\ data cube and
checked the probability of an OH line that happens to overlay a CO line ($P_2$).

We identified clumps in the 1720~MHz OH data cube in each velocity channel from $-58.5$
to $135~\km\ps$
by applying the FellWalker clumpfind algorithm \citep{berry15} in the STARLINK package. 
The search criteria include at least 12 pixels in a single channel, with over 2$\sigma$ detection. 
These criteria are similar to those in \cite{beuther19}, but we identified spots in the single channel for narrow lines, searched for clumps with larger significance levels and used two tiles of data for cross-matching.
\cite{beuther19} show that no masers are detected in over two consecutive velocity channels in this sky region.
This over two-channel criterion is used to reduce spurious detection, but it could omit narrow masers.

\begin{figure}[h]
  \centering
\includegraphics[width=0.5\textwidth]{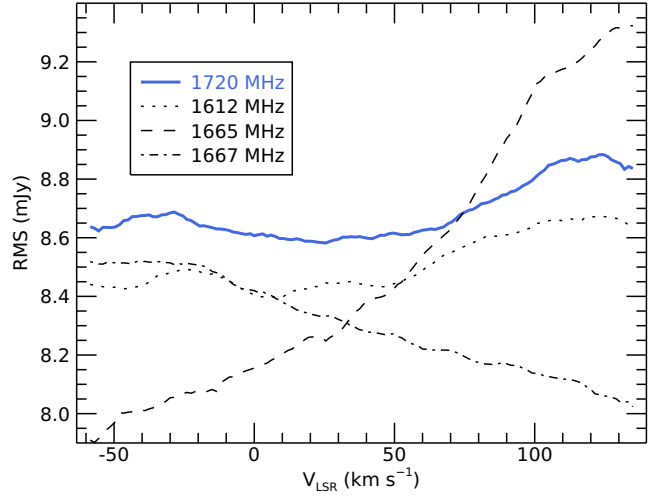}
\caption{
The channel-dependent rms of the 1720/1612/1665/1667~MHz OH data, averaged across the large-scale region. 
}
\label{fig:rms-chan}
\end{figure}

There is a spatial variation of the rms across the field, which influenced the significance level of the identified clumps and should be considered.
We calculated the rms in each pixel using all velocity channels and obtained the minimum, mean, and maximum rms values of 5.61~mJy, 8.78~mJy, and 12.5~mJy, respectively, for the tile at $l=56.^\circ{75}$.
We also found a variation of rms with the velocity channel (generally $<10\%$, but could be over 20\% in some pixels).
In general, the larger rms is found in the edge channels
for 1720~MHz OH data, especially channels with velocities $>80~\km\ps$ (see Figure~\ref{fig:rms-chan}), but the rms (channel) shape varies across the data.
Here we searched for maser spots in two rms cases. In Case A, we only consider the 
spatial rms variation. In Case B, we take the 
channel-dependent variation of rms into account. The rms in each channel was then calculated using the 40 neighboring channels. In the 20 edge channels, the rms might have larger uncertainties as more data were taken from one side.

\subsection{Case A}

\begin{figure}
\epsscale{1.1}
\plottwo{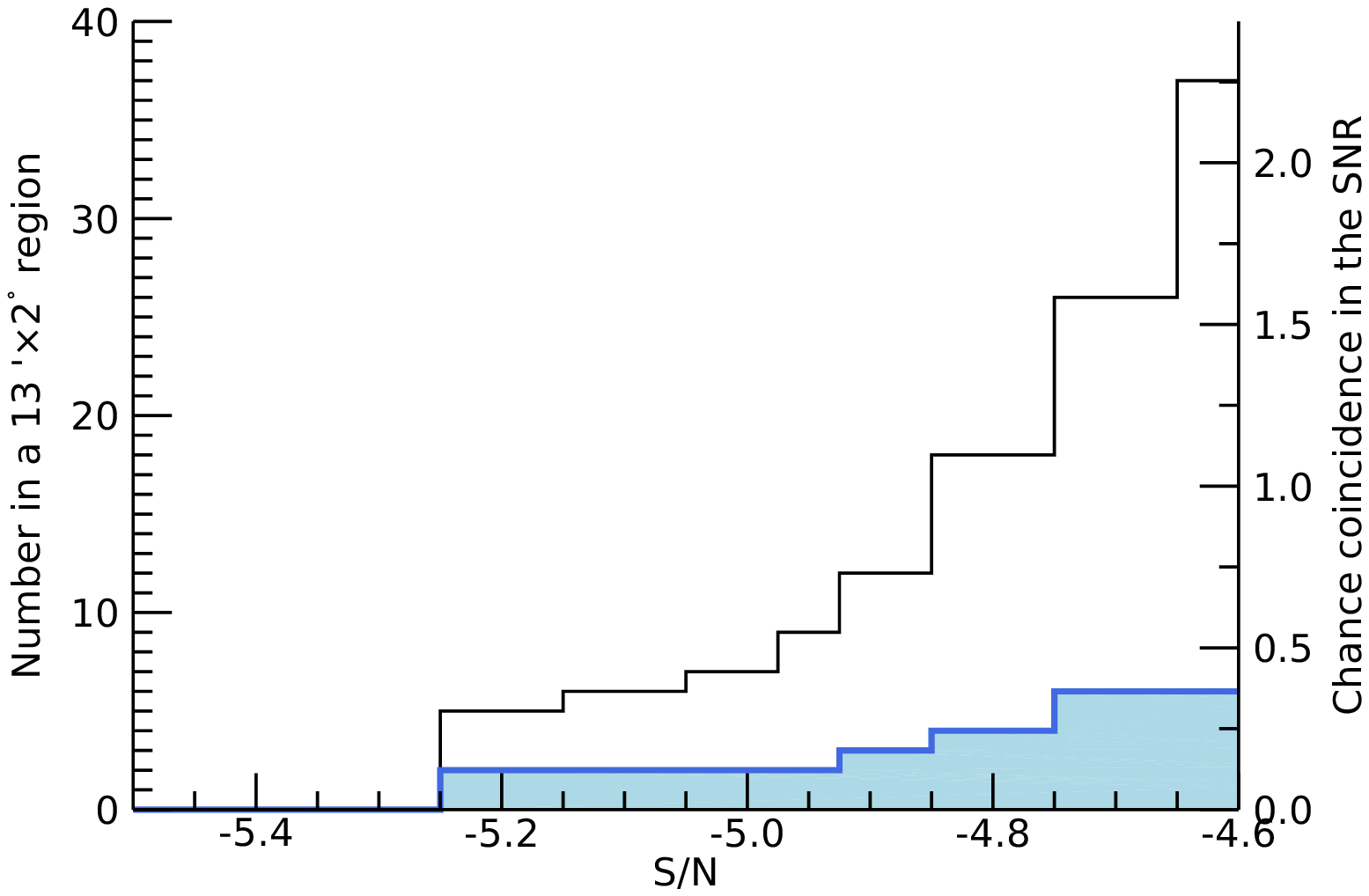}{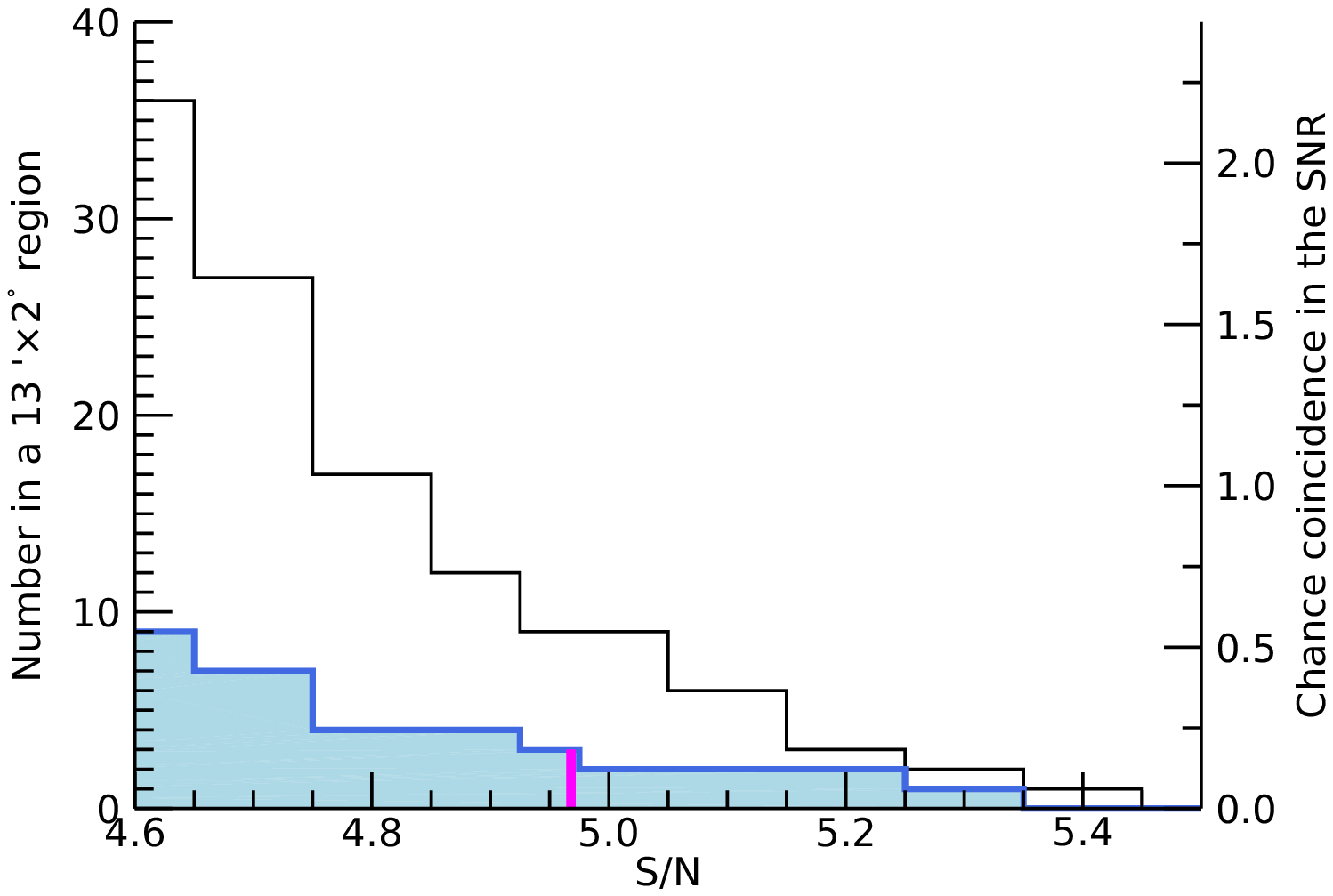}
\caption{
The identified 1720~MHz OH spots in a $13' \times 2^\circ$ region with S/N in the range of $-5.5$ to $-4.6$ (left panel) and 4.6 -- 5.5 (right panel), assuming that the noise is independent of the velocity channel (Case A).
The left and right axes give the detection number and the chance coincident rate in the SNR.
The black line shows the identified number in 
from the single tile $l=56.^\circ{75}$, which 
suffered extra noise due to the cleaning of the
interferometric data.
The blue line with the shaded region shows the identified spots in both tiles. 
The magenta vertical line denotes the lower limit of the S/N for the $\VLSR=30\km\ps$ maser spot in both tiles at $l=58^\circ$ and $56.^\circ 75$, respectively.
\label{fig:1720OH_detection}}
\end{figure}

Assuming that the noise is independent of the velocity channel,
we found all the detected spots with $|$S/N$|<5.4$.
We searched for clumps with peak values with signal-to-noise ratio (S/N)=4.5--6 and $-6$--$-4.5$.
The spots with negative S/N and flux are false detection and are used to test the noise distribution as a function of S/N.
Figure~\ref{fig:1720OH_detection} shows
the detection number in the $13'\times 2^\circ$ region and the fraction of the identified clumps
in the SNR given the area ratio.
The black line shows the results from the tile at $l=56.^\circ{75}$ alone, while the blue shaded area shows the cross-matched detection using both tiles at $l=56.^\circ{75}$ and $l=58^\circ$.
The figure proves that over 3/4 of the identified spots with S/N $< 5$ are false signals due to different cleaning areas.
None of the S/N$\ge 4.7$ spots have 1612/1665/1667 OH counterparts. 
The two panels in Figure~\ref{fig:1720OH_detection} show that
positive and negative noise distribution is nearly symmetric.
They show that $|$S/N$|\sim 5$ approaches the end of the noise tail.

At the significance of 4.95, we identified three spots in the large region ($16.4$ times the SNR area), including the $\VLSR=30\km\ps$ spot at the boundary of \snr.
For comparison, there are two confirmed noise spots with S/N$<-4.95$. 
Assuming that all the identified spots are false signals, 
for randomly finding one spot with a peak significance of 4.95 in both tiles in an area as large as the SNR, the probability is only 18\% (12\% when considering the spots with negative values).

As OH is a molecule, the real OH line is expected to correspond to molecular emission at around the same LSR velocity.
In contrast,
the false detection would have a randomly distributed $\VLSR$ 
and does not need a \twCO\ counterpart.
At the position of 1720~MHz OH maser near \snr,
only $\sim 7$\% of channels have over 3-$\sigma$ \twCO\ emission in 
$\VLSR=-58.5$--$135~\km\ps$.
We retrieved PMO \twCO~\Jotz\ data for a large sky region centered at $l=57.^\circ 235$, $b=0.68^\circ$ and with a size of $13'\times 1^\circ$.
We found that  the over 3-$\sigma$ \twCO\ line is shown 
in $<22\%$ (mean value of 6.7\%) of the velocity channels.
This means the probability for an OH--\twCO\ coincidence is low and can be used to
filter out false detection.

At the 4.95$\sigma$ level, the probability of an OH spot randomly falling in the SNR is $P_1= 18\%$ and the probability of OH--\twCO\ correspondence is $P_2=7\%$.
As $P_1$ and $P_2$ are independent, the false detection rate is $P_f=P_1 P_2\sim 1\%$ for the 1720 MHz OH spot that
we identified at the boundary of \snr.
Therefore, the probability of the real detection
of the 1720~MHz maser is roughly estimated as $P=1-P_f\sim 99\%$.
Nevertheless, the next two subsections will show that the channel-dependent 
rms variation indeed changes the results and should be considered.

\begin{figure}
\epsscale{1.1}
\plottwo{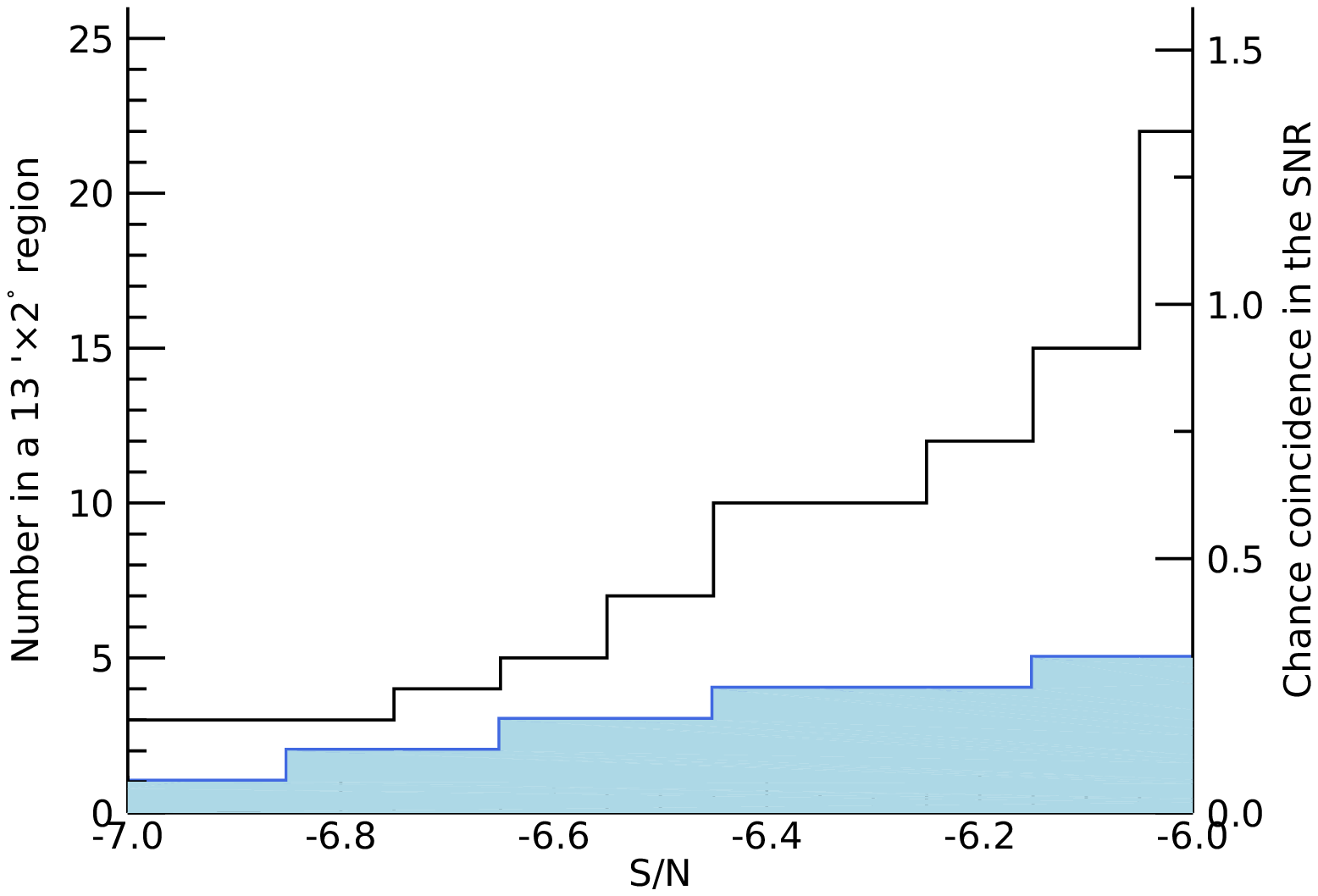}{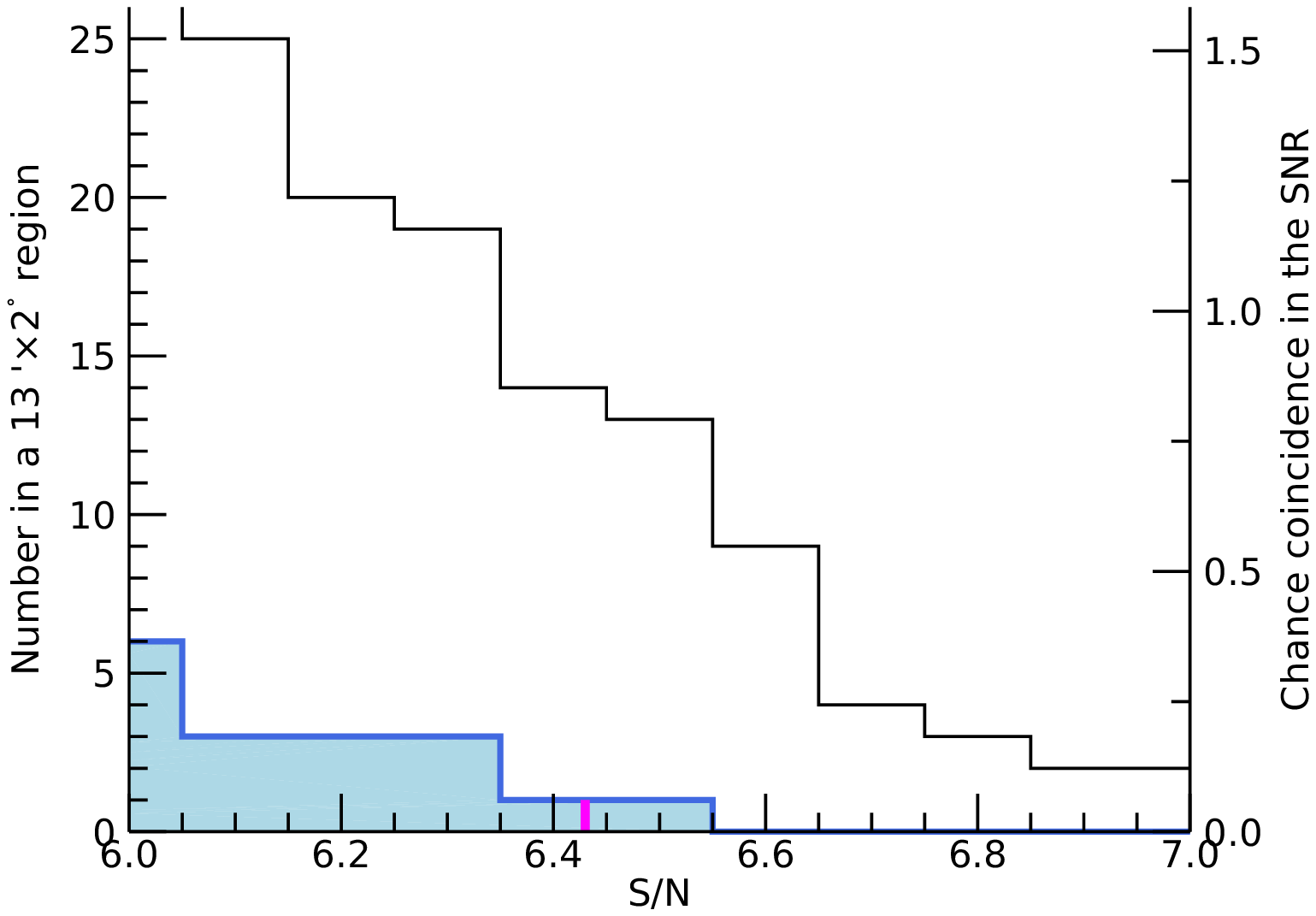}
\caption{
The figure is similar to Figure~\ref{fig:1720OH_detection}, but
the spot identification considers the varied noise across the velocity channels.
\label{fig:1720OH_detection_var}}
\end{figure}

\subsection{Case B}

We repeated the clump identification process, but including the variation
of rms with velocity channels. 
Figure~\ref{fig:1720OH_detection_var} shows the identified clumps. The $\VLSR=30\km\ps$ maser spot near \snr\ is identified
in both tiles and is the only spot with a minimum S/N$=6.4$.
The S/N of the spot is larger than in Case A, as the rms at $\VLSR=30~\km\ps$ is generally lower than the edge channels.
We noted that the noise distribution of
Case B is less symmetric than Case A and the negative noise tail extends to the larger $|$S/N$|$.
Nevertheless, in Case B, the S/N of our maser spot is still at the end of the noise tail, and the chance coincidence in the SNR is still low ($P_1=6\%$).
In this case, the probability of the real detection of 1720~MHz maser in the SNR is
$P=1-P_1 P_2=99.6\%$.

\subsection{Spot identification in 1612/1665/1667 MHz OH data}

\begin{figure*}[h]
  \centering
\includegraphics[width=0.3\textwidth]{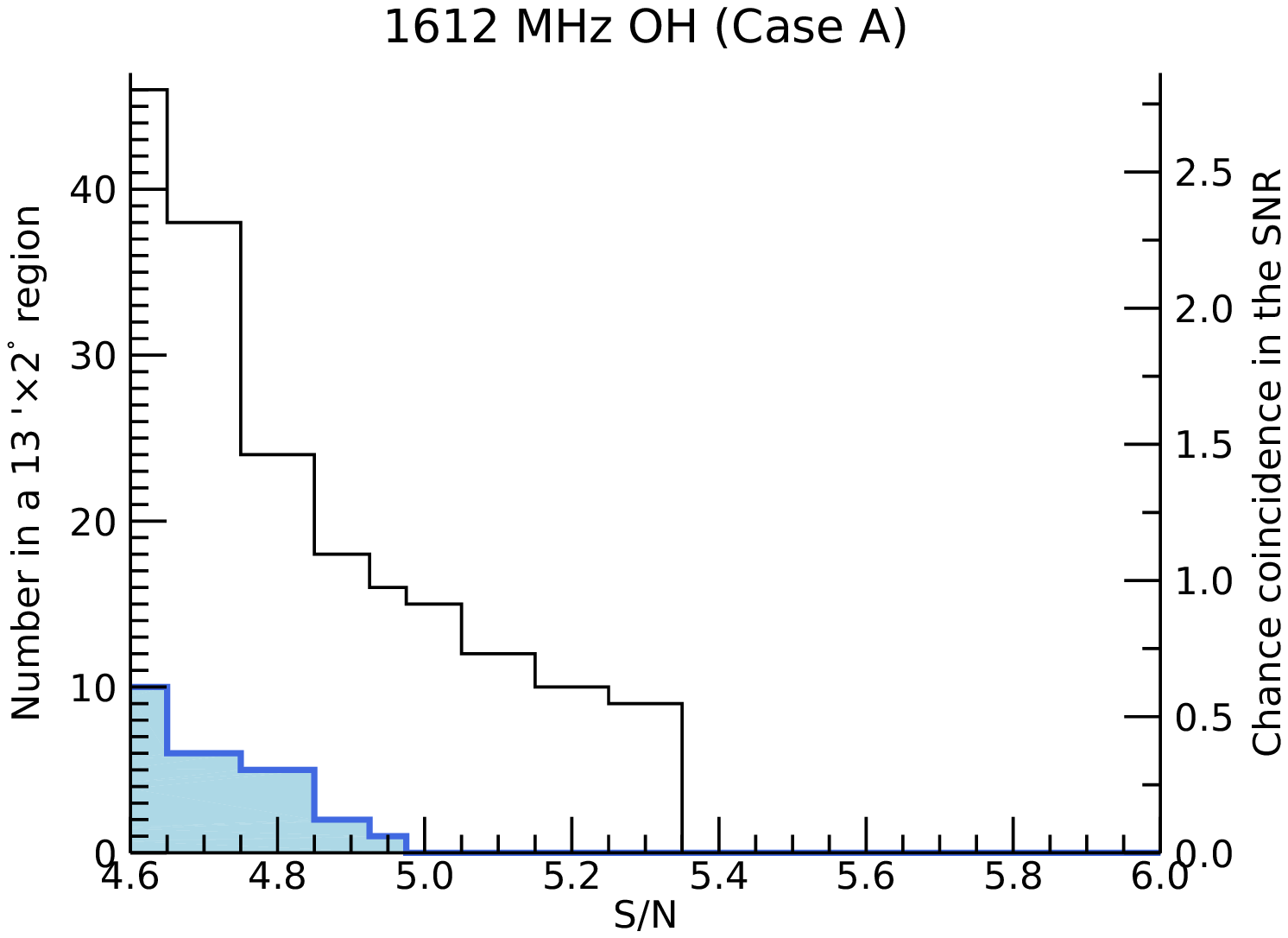}
\includegraphics[width=0.3\textwidth]{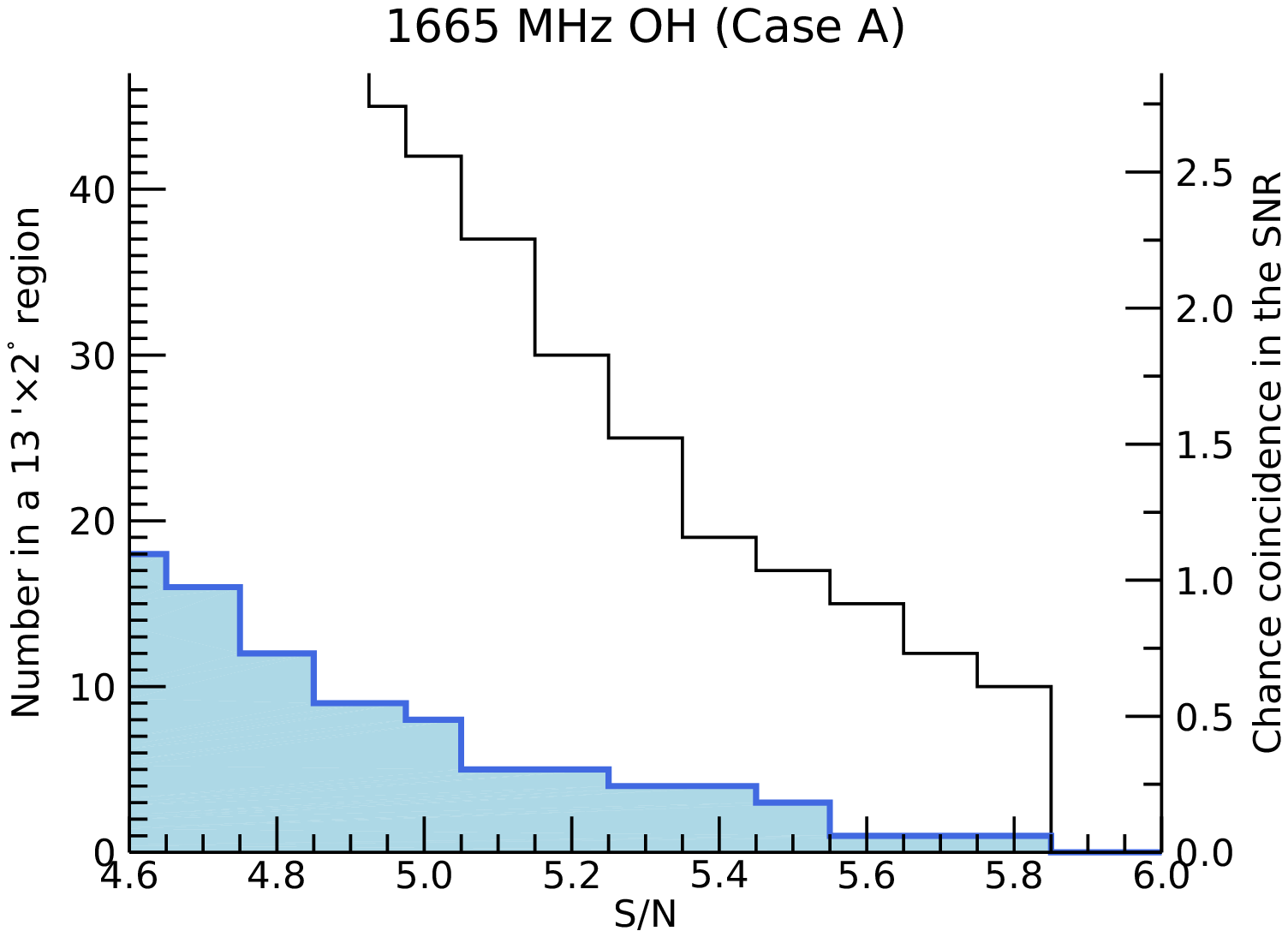}
\includegraphics[width=0.3\textwidth]{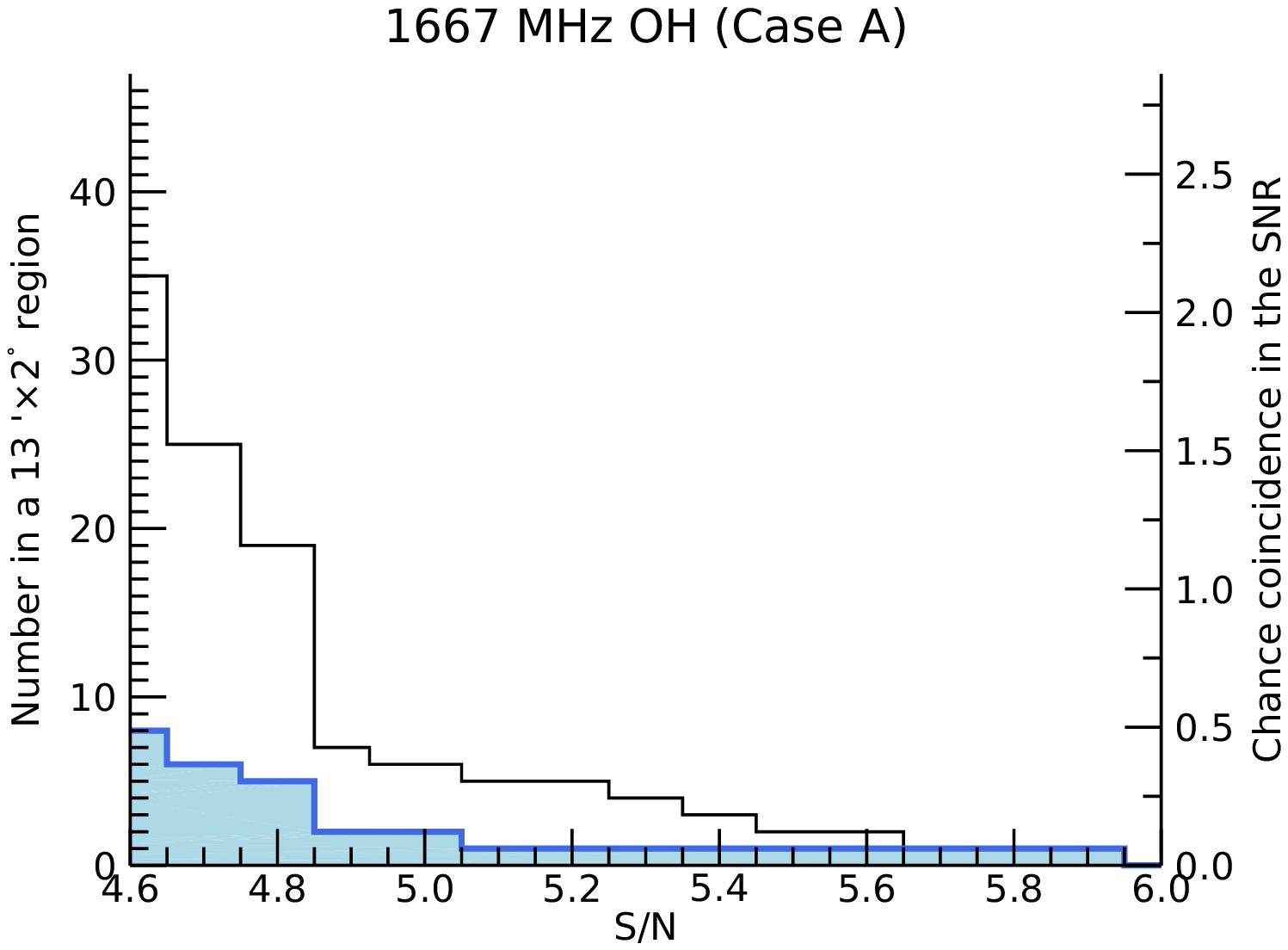}
\includegraphics[width=0.3\textwidth]{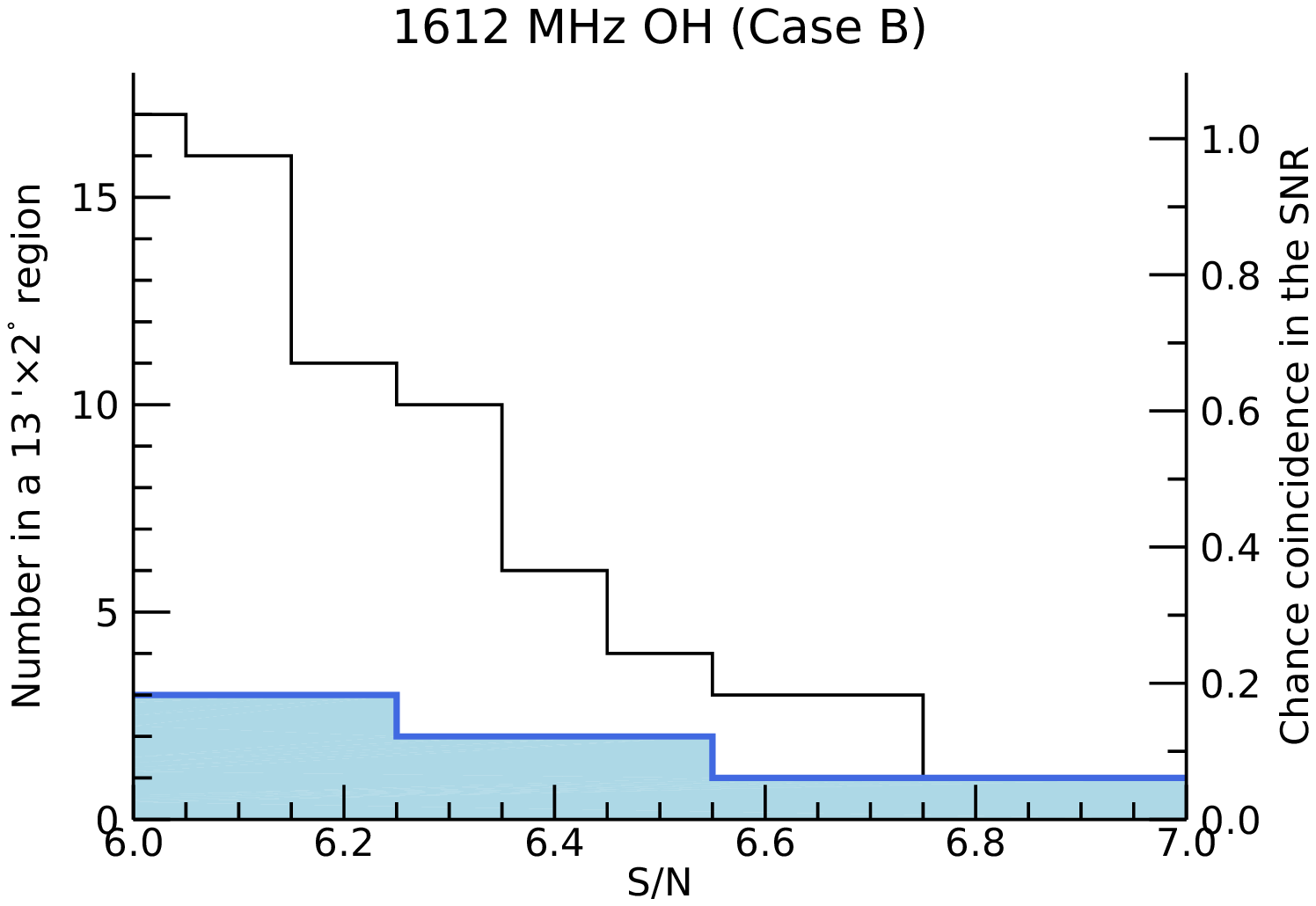}
\includegraphics[width=0.3\textwidth]{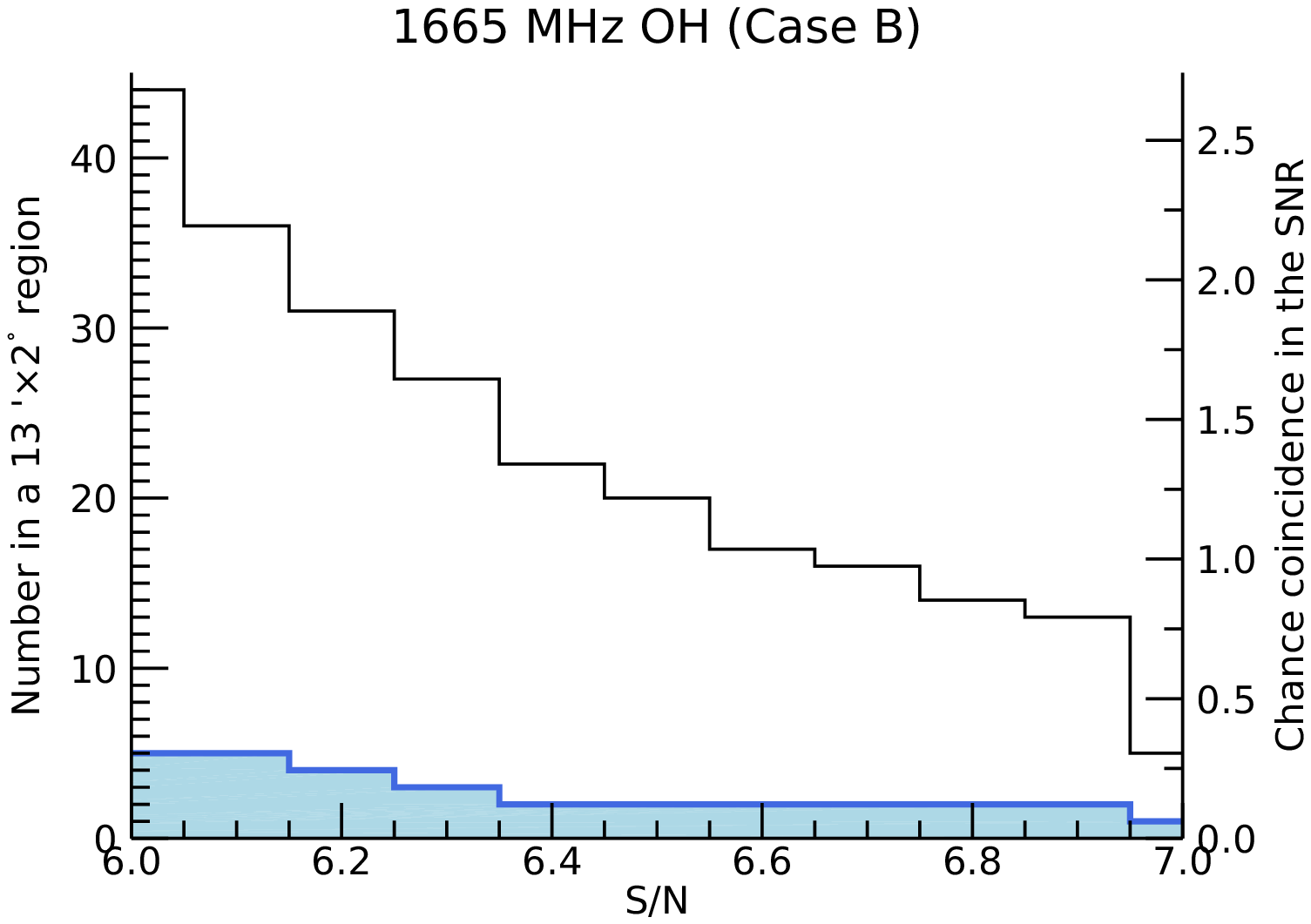}
\includegraphics[width=0.3\textwidth]{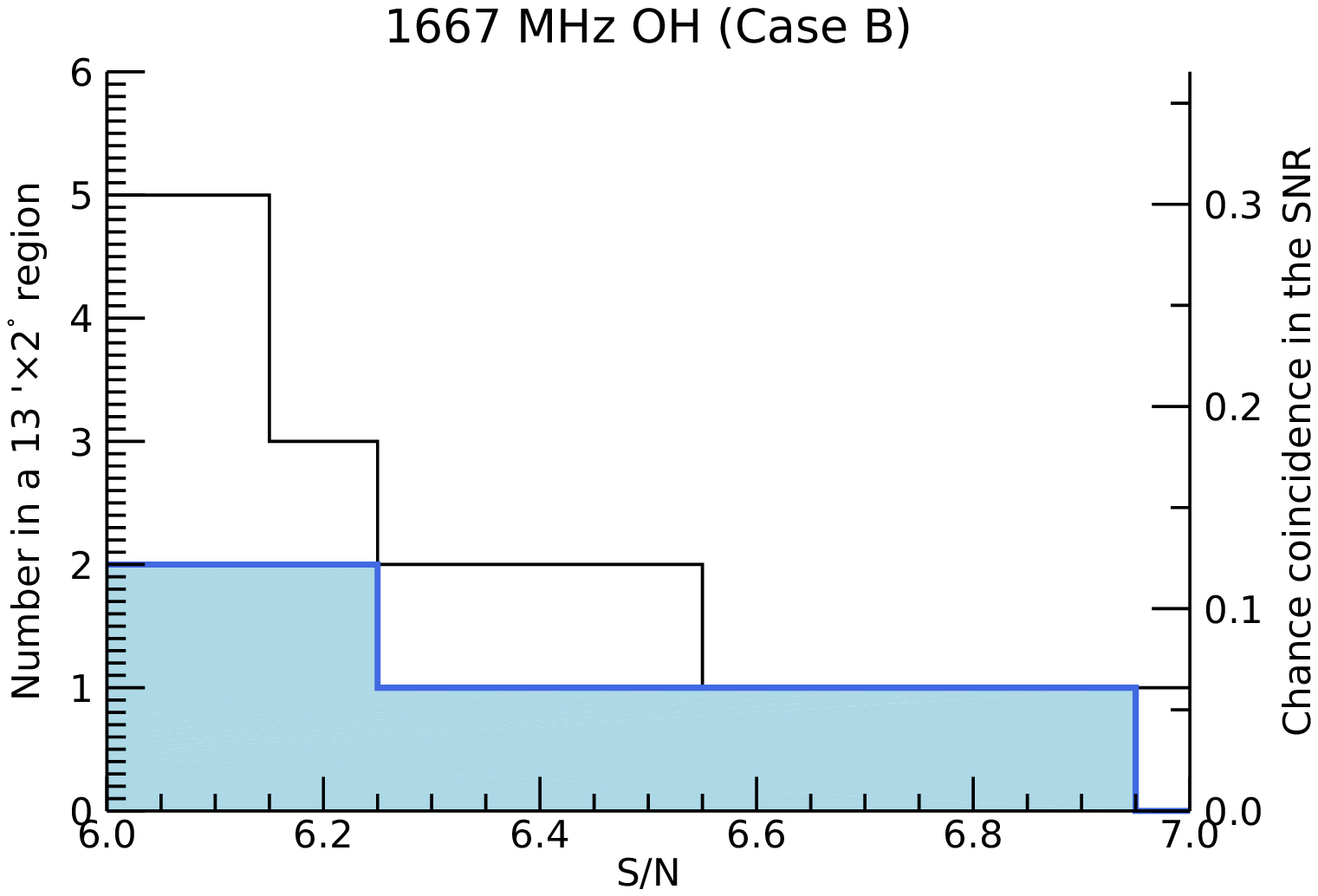}
\caption{
The images are similar to the right panels of Figures~\ref{fig:1720OH_detection} and \ref{fig:1720OH_detection_var}, but for three other
OH transitions. 
}
\label{fig:otherOH_identification}
\end{figure*}

For comparison purposes, we searched for OH clumps using
1612, 1665, and 1667 MHz THOR data and checked
their S/N distributions.
The identification processes are the same as that for
the 1720~MHz OH data.
Figure~\ref{fig:otherOH_identification} shows the identified positive-flux spots as a function of the S/N.
For Case A, at the significance of 4.95, the chance
coincidence in the SNR $P_1$ is 6\%, 55\%, 12\% for 1612, 1665, and 1667 MHz OH data, respectively.
The 1665 MHz data have a significantly larger detection than other data, due to the channel-dependent rms distribution.
Eight in nine spots are at $\VLSR> 88$~\kms, where the rms is larger than the mean rms level (see Figure~\ref{fig:rms-chan}).
This favors Case B as a meaningful case.

In Case B,  at the significance of 6.4, the chance coincidence for a random spot found in the SNR $P_1$ is 12\%, 12\% and 6\% for 1612, 1665, and 1667 MHz data for
the full channels.
The $P_1$ values are similar to the number of 6\% for 1720 MHz data.
All of the identified spots at around edge channels with the LSR
velocities larger than $88~\km\ps$ or smaller than $-42~\km\ps$, where the rms is enhanced (see Figure~\ref{fig:rms-chan}).

The comparison between Cases A and B shows that the spots
tend to be identified at the channels with underestimated rms.
If we exclude the edge channels with high rms and only consider the velocities in $-42$--88~\kms, the 1720~MHz
OH spot near \snr\ is the only identified spot with S/N$>6.4$ among all THOR OH data.
We verify that only one spot among the four 1720 MHz OH spots with S/N$<-6.4$ falls in this
velocity range.
This reinforces a larger chance to find spots in these edge channels.

We conservatively set the chance coincidence $P_1$ to 12\%, using the upper limit in four groups of OH data and considering all channels, although the false rate should be much lower as all the other spots are detected at round edge channels.

The above analysis 
provides evidence that the 1720~MHz OH maser spot near \snr\ is not a random OH noise that happens to correspond to \twCO\ emission. 
Although it is faint, the false possibility is only $< 1\%$ ($P_1 P_2$).
Actually, faint and narrow 1720~MHz OH masers have also been found in other SNRs.
\cite{hewitt09} detected two faint OH masers with a peak flux less than 70~mJy beam$^{-1}$.
Two faint 1720~MHz OH masers were detected in N49 \citep{brogan04}, an SNR hosting magnetar SGR~0526$-66$. 
The brighter maser in N49 has a peak flux density of $35$ mJy beam$^{-1}$ and has a line width less than $1 \km\ps$, even weaker than that reported here.

\begin{figure}
\epsscale{0.5}
\plotone{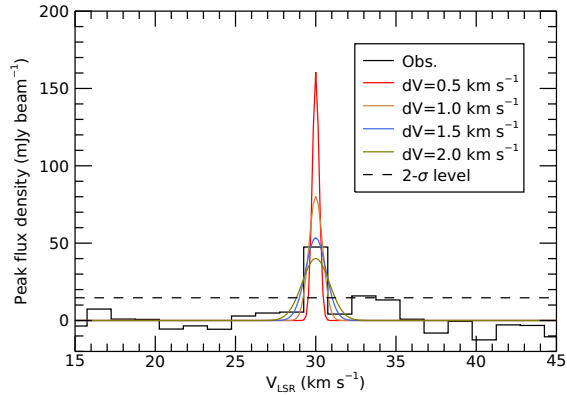}
\caption{
A comparison of the observed 1720 MHz OH maser and a few Gaussian lines with different line widths but the same area as that of the observed line.
\label{fig:maser_gaufit}}
\end{figure}

\section{Line widths of the OH maser}

The OH maser near \snr\ is only detected in a single channel due to the
narrow line width.
Here we examined whether the line wing could be bright enough to be detected in 
the adjacent channels if the line peak is at the channel center.
Figure~\ref{fig:oh} shows that the observed line wing of the maser is at $< 1\sigma$ level, below the detection level.
Considering the line peak is at 6.4$\sigma$ level, we expect
to detect $> 76\%$ flux in the central channel.
According to the Gauss error function, the flux fraction is 
${\rm erf}(z/\sqrt{2})> 0.76$, which gives $z < 1.175$. 
The value $z$ can be expressed using the channel resolution $dV_c=1.5~\km\ps$ and the FWHM of the line $dV$:
$z=\sqrt{2 \ln{2}} dV_c/dV $.
Therefore, we obtained and $dV < 1.5\km\ps$. 
Setting the wing-channel significance at $2\sigma$ will give $z\le 0.87$ and a
line width $dV \le 2~\km\ps$.
The line centroid mismatches the channel center, then the line width should to be lower than the aforementioned values.

Figure~\ref{fig:maser_gaufit} shows the observed spectrum and four exemplified Gaussian
lines with a velocity-integrated flux density equal in the area of the three channels ($85.6$~mJy beam$^{-1}$ \kms).
The figure displays that lines with widths $dV \le 1.5 \km\ps$ may not be detected in more than one channel, as the line wings are too faint.

\section{HI morphology at $\VLSR=29.5~\km\ps$} \label{sec:HI}

\begin{figure*}[h]
  \centering
\includegraphics[width=0.48\textwidth]{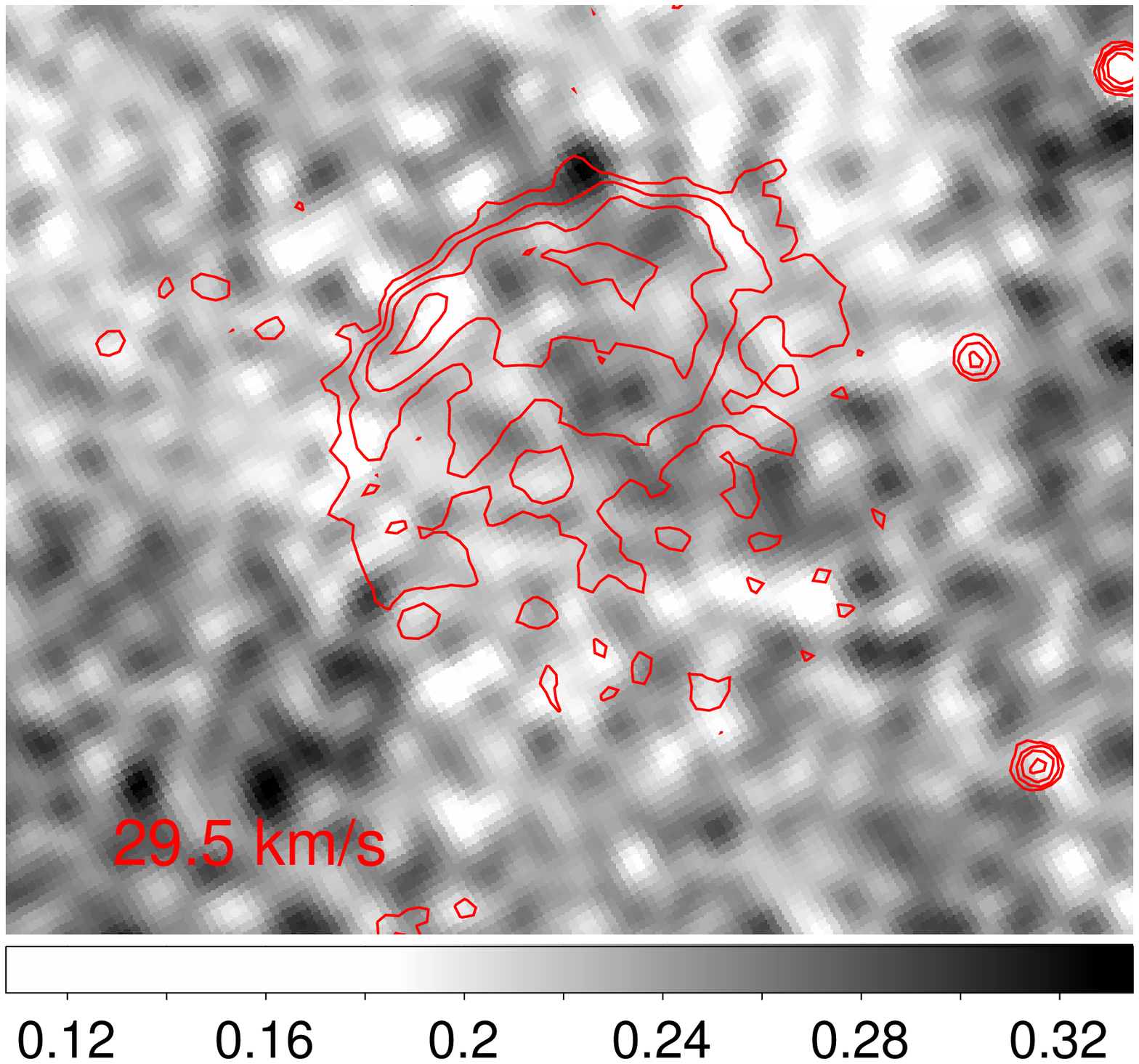}
\includegraphics[width=0.48\textwidth]{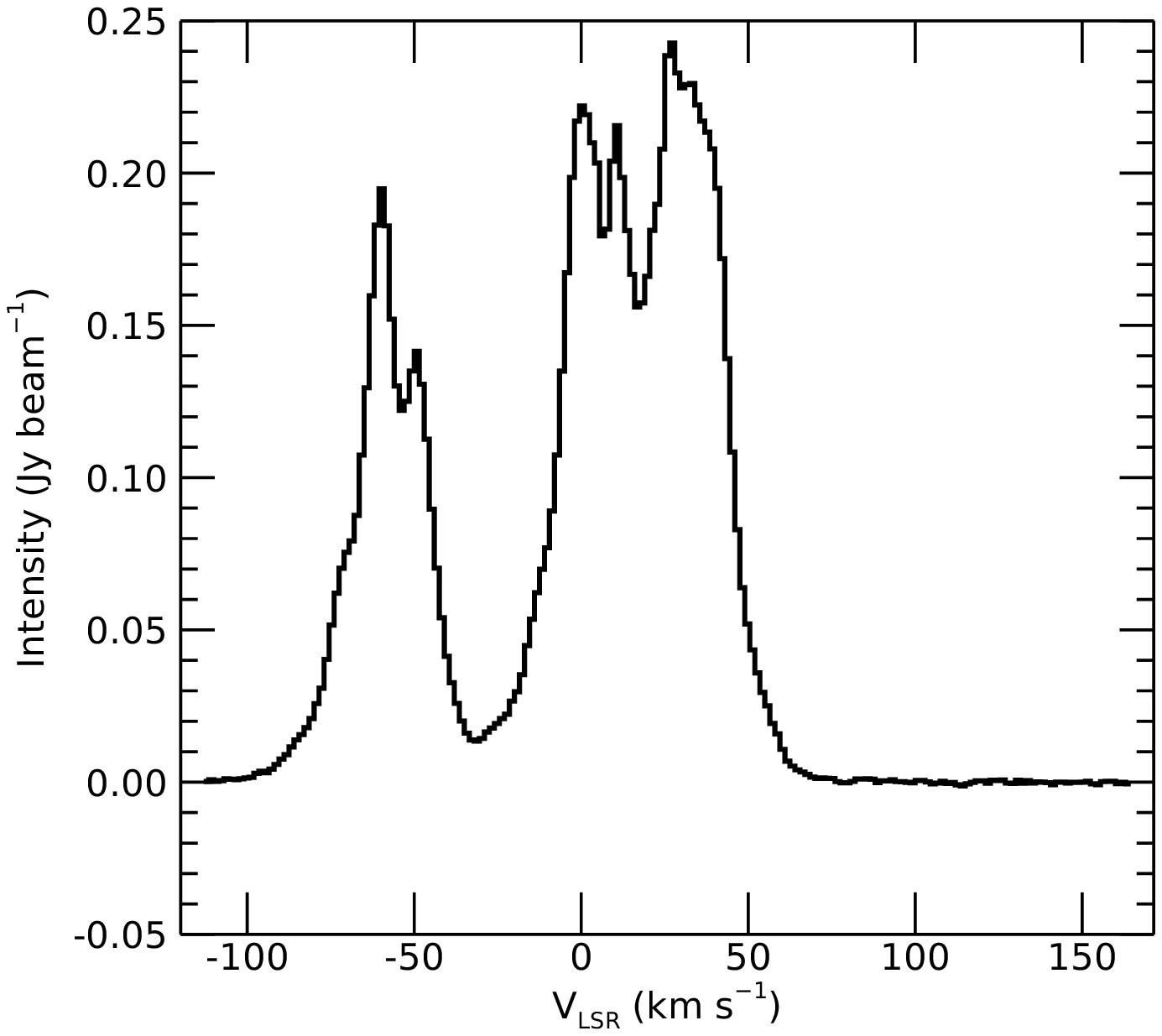}
\caption{
Left: HI morphology at 29.5~$\km\ps$. The unit of the color bars is Jy beam$^{-1}$.
Right: Averaged HI spectrum extracted from the SNR with a radius of $5\farcm{5}$.
}
\label{fig:HI}
\end{figure*}

\begin{figure}
\epsscale{1.0}
\plotone{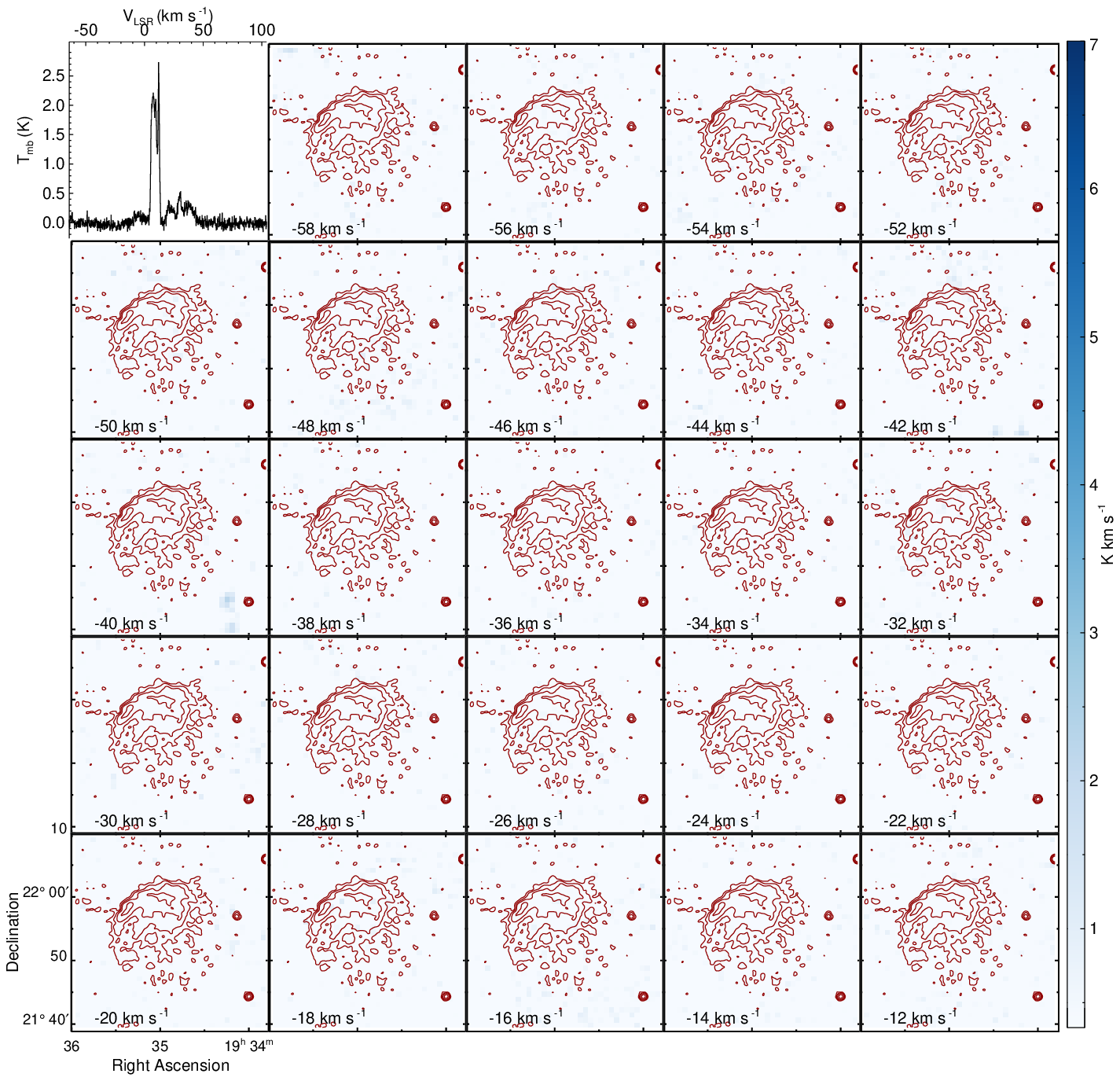}
\caption{
A grid of the velocity-integrated intensity maps of PMO \twCO~\Jotz\ emission in $\VLSR=-59$ to $-11~\km\ps$ with a velocity step of 2~\kms.
The contours are taken from the THOR 1.4~GHz radio continuum.
The first panel shows the \twCO~\Jotz\ spectrum averaged across
the field-of-view.
\label{fig:cogrid_negv}}
\end{figure}

As shown in Figure~\ref{fig:HI}, the 
HI morphology at $\VLSR=29.5~\km\ps$ is relatively smooth, with a bright knot in the SNR north.
%There is no clear correspondence between the SNR and an HI shell at $\sim 30~\km\ps$.
There are two possible explanations for the lack of structured HI emission at $\VLSR\sim 30~\km\ps$.
The first possible reason is that \snr\ is not in an 
atomic environment.
So far, the HI studies of \snr\ have not provided any kinematic evidence to support
that SNR \snr\ perturbs or heats HI gas \citep[see][]{surnis16,kothes18,ranasinghe18}. 
Although a morphological match between an HI structure at $\VLSR=-46~\km\ps$ and the SNR has been proposed \citep{kothes18}, the chance coincidence has not been ruled out.
We have not found relevant CO structures at $\VLSR=-46~\km\ps$
(see Figure~\ref{fig:cogrid_negv}).
In the alternative scenario, the SNR is in a relatively homogeneous atomic environment, and it is difficult to identify a shocked HI structure.
The HI emission line is broad in the inner Galaxy, and the line crowding is severe. The right panel of Figure~\ref{fig:HI} shows that
there are multiple HI components from $\VLSR=-100~\km\ps$ to $60~\km\ps$ toward \snr.
The HI morphology in a single channel can be easily contaminated by HI wings of nearby velocity components. In this case, an HI enhancement might not be discerned even if the SNR heats the atomic gas. 
A detailed exploration of shocked HI emission is out of the scope of this paper.
From the HI images, we cannot conclude or refute a relationship between the
SNR and the atomic gas at $\VLSR\sim 30 \km\ps$.

\end{CJK*}

\begin{thebibliography}{}
\expandafter\ifx\csname natexlab\endcsname\relax\def\natexlab#1{#1}\fi
\providecommand{\url}[1]{\href{#1}{#1}}

\bibitem[{{Arikawa} {et~al.}(1999){Arikawa}, {Tatematsu}, {Sekimoto}, \&
  {Takahashi}}]{arikawa99}
{Arikawa}, Y., {Tatematsu}, K., {Sekimoto}, Y., \& {Takahashi}, T. 1999, \pasj,
  51, L7

\bibitem[{{Barthelmy} {et~al.}(2020){Barthelmy}, {Bernardini}, {D'Avanzo},
  {Gropp}, {Kennea}, {Lien}, {Meland ri}, {Palmer}, {Sbarrato}, {Siegel}, \&
  {Neil Gehrels Swift Observatory Team}}]{barthelmy20}
{Barthelmy}, S.~D., {Bernardini}, M.~G., {D'Avanzo}, P., {et~al.} 2020, GRB
  Coordinates Network, 27657, 1

\bibitem[{{Berry}(2015)}]{berry15}
{Berry}, D.~S. 2015, Astronomy and Computing, 10, 22

\bibitem[{{Beuther} {et~al.}(2016){Beuther}, {Bihr}, {Rugel}, {Johnston},
  {Wang}, {Walter}, {Brunthaler}, {Walsh}, {Ott}, {Stil}, {Henning},
  {Schierhuber}, {Kainulainen}, {Heyer}, {Goldsmith}, {Anderson}, {Longmore},
  {Klessen}, {Glover}, {Urquhart}, {Plume}, {Ragan}, {Schneider},
  {McClure-Griffiths}, {Menten}, {Smith}, {Roy}, {Shanahan}, {Nguyen-Luong}, \&
  {Bigiel}}]{beuther16}
{Beuther}, H., {Bihr}, S., {Rugel}, M., {et~al.} 2016, \aap, 595, A32

\bibitem[{{Beuther} {et~al.}(2019){Beuther}, {Walsh}, {Wang}, {Rugel}, {Soler},
  {Linz}, {Klessen}, {Anderson}, {Urquhart}, {Glover}, {Billington},
  {Kainulainen}, {Menten}, {Roy}, {Longmore}, \& {Bigiel}}]{beuther19}
{Beuther}, H., {Walsh}, A., {Wang}, Y., {et~al.} 2019, \aap, 628, A90

\bibitem[{{Bochenek} {et~al.}(2020){Bochenek}, {Ravi}, {Belov}, {Hallinan},
  {Kocz}, {Kulkarni}, \& {McKenna}}]{bochenek20}
{Bochenek}, C.~D., {Ravi}, V., {Belov}, K.~V., {et~al.} 2020, \nat, 587, 59

\bibitem[{{Braun} {et~al.}(2019){Braun}, {Safi-Harb}, \& {Fryer}}]{braun19}
{Braun}, C., {Safi-Harb}, S., \& {Fryer}, C.~L. 2019, \mnras, 489, 4444

\bibitem[{{Brogan} {et~al.}(2004){Brogan}, {Goss}, {Lazendic}, \&
  {Green}}]{brogan04}
{Brogan}, C.~L., {Goss}, W.~M., {Lazendic}, J.~S., \& {Green}, A.~J. 2004, \aj,
  128, 700

\bibitem[{{Chen} {et~al.}(2014){Chen}, {Jiang}, {Zhou}, {Su}, {Zhou}, {Li}, \&
  {Zhang}}]{chen14}
{Chen}, Y., {Jiang}, B., {Zhou}, P., {et~al.} 2014, in IAU Symposium, Vol. 296,
  Supernova Environmental Impacts, ed. A.~{Ray} \& R.~A. {McCray}, 170--177

\bibitem[{{Chen} {et~al.}(2013){Chen}, {Zhou}, \& {Chu}}]{chen13}
{Chen}, Y., {Zhou}, P., \& {Chu}, Y.-H. 2013, \apjl, 769, L16

\bibitem[{{Chevalier}(1974)}]{chevalier74}
{Chevalier}, R.~A. 1974, \apj, 188, 501

\bibitem[{{Chevalier}(1999)}]{chevalier99}
---. 1999, \apj, 511, 798

\bibitem[{{Cioffi} {et~al.}(1988){Cioffi}, {McKee}, \&
  {Bertschinger}}]{cioffi88}
{Cioffi}, D.~F., {McKee}, C.~F., \& {Bertschinger}, E. 1988, \apj, 334, 252

\bibitem[{{Claussen} {et~al.}(1997){Claussen}, {Frail}, {Goss}, \&
  {Gaume}}]{claussen97}
{Claussen}, M.~J., {Frail}, D.~A., {Goss}, W.~M., \& {Gaume}, R.~A. 1997, \apj,
  489, 143

\bibitem[{{Cummmings} {et~al.}(2014){Cummmings}, {Barthelmy}, {Chester}, \&
  {Page}}]{cummings14}
{Cummmings}, J.~R., {Barthelmy}, S.~D., {Chester}, M.~M., \& {Page}, K.~L.
  2014, The Astronomer's Telegram, 6294, 1

\bibitem[{{Currie} {et~al.}(2014){Currie}, {Berry}, {Jenness}, {Gibb}, {Bell},
  \& {Draper}}]{currie14}
{Currie}, M.~J., {Berry}, D.~S., {Jenness}, T., {et~al.} 2014, in Astronomical
  Society of the Pacific Conference Series, Vol. 485, Astronomical Data
  Analysis Software and Systems XXIII, ed. N.~{Manset} \& P.~{Forshay}, 391

\bibitem[{{Draine}(2011)}]{drain11}
{Draine}, B.~T. 2011, {Physics of the Interstellar and Intergalactic Medium}

\bibitem[{{Elitzur}(1992)}]{elitzur92}
{Elitzur}, M. 1992, \araa, 30, 75

\bibitem[{{Fletcher} \& {Fermi GBM Team}(2020)}]{fletcher20}
{Fletcher}, C., \& {Fermi GBM Team}. 2020, GRB Coordinates Network, 27659, 1

\bibitem[{{Frail} {et~al.}(1996){Frail}, {Goss}, {Reynoso}, {Giacani}, {Green},
  \& {Otrupcek}}]{frail96}
{Frail}, D.~A., {Goss}, W.~M., {Reynoso}, E.~M., {et~al.} 1996, \aj, 111, 1651

\bibitem[{{Frail} {et~al.}(1994){Frail}, {Goss}, \& {Slysh}}]{frail94}
{Frail}, D.~A., {Goss}, W.~M., \& {Slysh}, V.~I. 1994, \apjl, 424, L111

\bibitem[{{Frail} \& {Mitchell}(1998)}]{frail98}
{Frail}, D.~A., \& {Mitchell}, G.~F. 1998, \apj, 508, 690

\bibitem[{{Gaensler}(2014)}]{gaensler14}
{Gaensler}, B.~M. 2014, GRB Coordinates Network, 16533, 1

\bibitem[{{Hewitt} {et~al.}(2009){Hewitt}, {Yusef-Zadeh}, \&
  {Wardle}}]{hewitt09}
{Hewitt}, J.~W., {Yusef-Zadeh}, F., \& {Wardle}, M. 2009, \apjl, 706, L270

\bibitem[{{Hollenbach} {et~al.}(2013){Hollenbach}, {Elitzur}, \&
  {McKee}}]{hollenbach13}
{Hollenbach}, D., {Elitzur}, M., \& {McKee}, C.~F. 2013, \apj, 773, 70

\bibitem[{{Israel} {et~al.}(2016){Israel}, {Esposito}, {Rea}, {Coti Zelati},
  {Tiengo}, {Campana}, {Mereghetti}, {Rodriguez Castillo}, {G{\"o}tz},
  {Burgay}, {Possenti}, {Zane}, {Turolla}, {Perna}, {Cannizzaro}, \&
  {Pons}}]{isarel16}
{Israel}, G.~L., {Esposito}, P., {Rea}, N., {et~al.} 2016, \mnras, 457, 3448

\bibitem[{{Jiang} {et~al.}(2010){Jiang}, {Chen}, {Wang}, {Su}, {Zhou},
  {Safi-Harb}, \& {DeLaney}}]{jiang10}
{Jiang}, B., {Chen}, Y., {Wang}, J., {et~al.} 2010, \apj, 712, 1147

\bibitem[{{Koo} {et~al.}(2016){Koo}, {Lee}, {Jeong}, {Seok}, \& {Kim}}]{koo16}
{Koo}, B.-C., {Lee}, J.-J., {Jeong}, I.-G., {Seok}, J.~Y., \& {Kim}, H.-J.
  2016, \apj, 821, 20

\bibitem[{{Kothes} {et~al.}(2018){Kothes}, {Sun}, {Gaensler}, \&
  {Reich}}]{kothes18}
{Kothes}, R., {Sun}, X., {Gaensler}, B., \& {Reich}, W. 2018, \apj, 852, 54

\bibitem[{{Kozlova} {et~al.}(2016){Kozlova}, {Israel}, {Svinkin}, {Frederiks},
  {Pal'shin}, {Tsvetkova}, {Hurley}, {Goldsten}, {Golovin}, {Mitrofanov}, \&
  {Zhang}}]{kozlova16}
{Kozlova}, A.~V., {Israel}, G.~L., {Svinkin}, D.~S., {et~al.} 2016, \mnras,
  460, 2008

\bibitem[{{Larson}(1981)}]{larson81}
{Larson}, R.~B. 1981, \mnras, 194, 809

\bibitem[{{Li} {et~al.}(2020){Li}, {Lin}, {Xiong}, {Ge}, {Li}, {Li}, {Lu},
  {Zhang}, {Tuo}, {Nang}, {Zhang}, {Xiao}, {Chen}, {Song}, {Xu}, {Liu}, {Jia},
  {Cao}, {Zhang}, {Qu}, {Liao}, {Zhao}, {Tan}, {Nie}, {Zhao}, {Zheng}, {Zheng},
  {Luo}, {Cai}, {Li}, {Xue}, {Bu}, {Chang}, {Chen}, {Chen}, {Chen}, {Chen},
  {Chen}, {Cui}, {Cui}, {Deng}, {Dong}, {Du}, {Fu}, {Gao}, {Gao}, {Gao}, {Gu},
  {Guan}, {Guo}, {Han}, {Huang}, {Huo}, {Jiang}, {Jiang}, {Jin}, {Jin}, {Kong},
  {Li}, {Li}, {Li}, {Li}, {Li}, {Li}, {Li}, {Liang}, {Liu}, {Liu}, {Liu},
  {Liu}, {Liu}, {Lu}, {Lu}, {Luo}, {Ma}, {Meng}, {Ou}, {Sai}, {Shang}, {Song},
  {Sun}, {Tao}, {Wang}, {Wang}, {Wang}, {Wang}, {Wang}, {Wen}, {Wu}, {Wu},
  {Wu}, {Xiao}, {Yang}, {Yang}, {Yang}, {Yang}, {Yi}, {Yin}, {You}, {Zhang},
  {Zhang}, {Zhang}, {Zhang}, {Zhang}, {Zhang}, {Zhang}, {Zhang}, {Zhang},
  {Zhang}, {Zhang}, {Zhang}, {Zhang}, {Zhang}, {Zhang}, {Zhang}, {Zhou},
  {Zhou}, {Zhu}, {Zhu}, \& {Zhuang}}]{li20}
{Li}, C.~K., {Lin}, L., {Xiong}, S.~L., {et~al.} 2020, arXiv e-prints,
  arXiv:2005.11071

\bibitem[{{Lockett} {et~al.}(1999){Lockett}, {Gauthier}, \&
  {Elitzur}}]{lockett99}
{Lockett}, P., {Gauthier}, E., \& {Elitzur}, M. 1999, \apj, 511, 235

\bibitem[{{Martin} {et~al.}(2014){Martin}, {Rea}, {Torres}, \&
  {Papitto}}]{martin14}
{Martin}, J., {Rea}, N., {Torres}, D.~F., \& {Papitto}, A. 2014, \mnras, 444,
  2910

\bibitem[{{McKee} \& {Cowie}(1975)}]{mckee75}
{McKee}, C.~F., \& {Cowie}, L.~L. 1975, \apj, 195, 715

\bibitem[{{Mereghetti} {et~al.}(2020){Mereghetti}, {Savchenko}, {Ferrigno},
  {G{\"o}tz}, {Rigoselli}, {Tiengo}, {Bazzano}, {Bozzo}, {Coleiro},
  {Courvoisier}, {Doyle}, {Goldwurm}, {Hanlon}, {Jourdain}, {von Kienlin},
  {Lutovinov}, {Martin-Carrillo}, {Molkov}, {Natalucci}, {Onori}, {Panessa},
  {Rodi}, {Rodriguez}, {S{\'a}nchez-Fern{\'a}ndez}, {Sunyaev}, \&
  {Ubertini}}]{mereghetti20}
{Mereghetti}, S., {Savchenko}, V., {Ferrigno}, C., {et~al.} 2020, \apjl, 898,
  L29

\bibitem[{{Olausen} \& {Kaspi}(2014)}]{olausen14}
{Olausen}, S.~A., \& {Kaspi}, V.~M. 2014, \apjs, 212, 6

\bibitem[{{Orlando} {et~al.}(2007){Orlando}, {Bocchino}, {Reale}, {Peres}, \&
  {Petruk}}]{orlando07}
{Orlando}, S., {Bocchino}, F., {Reale}, F., {Peres}, G., \& {Petruk}, O. 2007,
  \aap, 470, 927

\bibitem[{{Palmer} \& {BAT Team}(2020)}]{palmer20}
{Palmer}, D.~M., \& {BAT Team}. 2020, GRB Coordinates Network, 27665, 1

\bibitem[{{Park} {et~al.}(2013){Park}, {Koo}, {Gibson}, {Kang}, {Lane},
  {Douglas}, {Peek}, {Korpela}, {Heiles}, \& {Newton}}]{park13a}
{Park}, G., {Koo}, B.~C., {Gibson}, S.~J., {et~al.} 2013, \apj, 777, 14

\bibitem[{{Pavlovic} {et~al.}(2014){Pavlovic}, {Dobardzic}, {Vukotic}, \&
  {Urosevic}}]{pavlovic14}
{Pavlovic}, M.~Z., {Dobardzic}, A., {Vukotic}, B., \& {Urosevic}, D. 2014,
  Serbian Astronomical Journal, 189, 25

\bibitem[{{Petroff} {et~al.}(2019){Petroff}, {Hessels}, \&
  {Lorimer}}]{petroff19}
{Petroff}, E., {Hessels}, J.~W.~T., \& {Lorimer}, D.~R. 2019, \aapr, 27, 4

\bibitem[{{Pinheiro Gon{\c{c}}alves} {et~al.}(2011){Pinheiro Gon{\c{c}}alves},
  {Noriega-Crespo}, {Paladini}, {Martin}, \& {Carey}}]{pinheirogoncalves11}
{Pinheiro Gon{\c{c}}alves}, D., {Noriega-Crespo}, A., {Paladini}, R., {Martin},
  P.~G., \& {Carey}, S.~J. 2011, \aj, 142, 47

\bibitem[{{Ranasinghe} {et~al.}(2018){Ranasinghe}, {Leahy}, \&
  {Tian}}]{ranasinghe18}
{Ranasinghe}, S., {Leahy}, D.~A., \& {Tian}, W. 2018, Open Physics Journal, 4,
  1

\bibitem[{{Reid} {et~al.}(2014){Reid}, {Menten}, {Brunthaler}, {Zheng}, {Dame},
  {Xu}, {Wu}, {Zhang}, {Sanna}, {Sato}, {Hachisuka}, {Choi}, {Immer},
  {Moscadelli}, {Rygl}, \& {Bartkiewicz}}]{reid14}
{Reid}, M.~J., {Menten}, K.~M., {Brunthaler}, A., {et~al.} 2014, \apj, 783, 130

\bibitem[{{Ridnaia} {et~al.}(2020){Ridnaia}, {Golenetskii}, {Aptekar},
  {Frederiks}, {Ulanov}, {Svinkin}, {Tsvetkova}, {Lysenko}, {Cline}, \&
  {Konus-Wind Team}}]{ridnaia20}
{Ridnaia}, A., {Golenetskii}, S., {Aptekar}, R., {et~al.} 2020, GRB Coordinates
  Network, 27669, 1

\bibitem[{{Shan} {et~al.}(2012){Shan}, {Yang}, {Shi}, {Yao}, {Zuo}, {Lin},
  {Chen}, {Zhang}, {Duan}, {Cao}, {Li}, {Li}, {Liu}, \& {Zhong}}]{shan12}
{Shan}, W., {Yang}, J., {Shi}, S., {et~al.} 2012, IEEE Transactions on
  Terahertz Science and Technology, 2, 593

\bibitem[{{Sieber} \& {Seiradakis}(1984)}]{sieber84}
{Sieber}, W., \& {Seiradakis}, J.~H. 1984, \aap, 130, 257

\bibitem[{{Surnis} {et~al.}(2016){Surnis}, {Joshi}, {Maan}, {Krishnakumar},
  {Manoharan}, \& {Naidu}}]{surnis16}
{Surnis}, M.~P., {Joshi}, B.~C., {Maan}, Y., {et~al.} 2016, \apj, 826, 184

\bibitem[{{Tavani} {et~al.}(2020){Tavani}, {Ursi}, {Verrecchia}, {Casentini},
  {Pittori}, {Pilia}, {Cardillo}, {Piano}, {Bulgarelli}, {Fioretti},
  {Parmiggiani}, {Lucarelli}, {Donnarumma}, {Vercellone}, {Gianotti},
  {Trifoglio}, {Giuliani}, {Mereghetti}, {Caraveo}, {Perotti}, {Chen}, {Argan},
  {Costa}, {Del Monte}, {Evangelista}, {Feroci}, {Lazzarotto}, {Lapshov},
  {Pacciani}, {Soffitta}, {Vittorini}, {Di Cocco}, {Fuschino}, {Galli},
  {Labanti}, {Marisaldi}, {Pellizzoni}, {Trois}, {Barbiellini}, {Vallazza},
  {Longo}, {Morselli}, {Picozza}, {Prest}, {Lipari}, {Zanello}, {Cattaneo},
  {Rappoldi}, {Ferrari}, {Paoletti}, {Antonelli}, {Giommi}, {Salotti},
  {Valentini}, \& {D'Amico}}]{tavani20}
{Tavani}, M., {Ursi}, A., {Verrecchia}, F., {et~al.} 2020, The Astronomer's
  Telegram, 13686, 1

\bibitem[{{The Chime/Frb Collaboration} {et~al.}(2020){The Chime/Frb
  Collaboration}, {Bandura}, {Bhardwaj}, {Bij}, {Boyce}, {Boyle}, {Brar},
  {Cassanelli}, {Chawla}, {Chen}, {Cliche}, {Cook}, {Cubranic}, {Curtin},
  {Denman}, {Dobbs}, {Dong}, {Fandino}, {Fonseca}, {Gaensler}, {Giri}, {Good},
  {Halpern}, {Hill}, {Hinshaw}, {H{\"o}fer}, {Josephy}, {Kania}, {Kaspi},
  {Landecker}, {Leung}, {Li}, {Lin}, {Masui}, {McKinven}, {Mena-Parra},
  {Merryfield}, {Meyers}, {Michilli}, {Milutinovic}, {Mirhosseini},
  {M{\"u}nchmeyer}, {Naidu}, {Newburgh}, {Ng}, {Patel}, {Pen},
  {Pinsonneault-Marotte}, {Pleunis}, {Quine}, {Rafiei-Ravandi}, {Rahman},
  {Ransom}, {Renard}, {Sanghavi}, {Scholz}, {Shaw}, {Shin}, {Siegel}, {Singh},
  {Smegal}, {Smith}, {Stairs}, {Tan}, {Tendulkar}, {Tretyakov}, {Vanderlinde},
  {Wang}, {Wulf}, \& {Zwaniga}}]{chime20}
{The Chime/Frb Collaboration}, Andersen, B.~C., {Bandura}, K.~M., {Bhardwaj},
  M., {et~al.} 2020, \nat, 587, 54

\bibitem[{{Vink}(2012)}]{vink12}
{Vink}, J. 2012, \aapr, 20, 49

\bibitem[{{Vink} \& {Kuiper}(2006)}]{vink06c}
{Vink}, J., \& {Kuiper}, L. 2006, \mnras, 370, L14

\bibitem[{{Wang} {et~al.}(2020){Wang}, {Beuther}, {Rugel}, {Soler}, {Stil},
  {Ott}, {Bihr}, {McClure-Griffiths}, {Anderson}, {Klessen}, {Goldsmith},
  {Roy}, {Glover}, {Urquhart}, {Heyer}, {Linz}, {Smith}, {Bigiel}, {Dempsey},
  \& {Henning}}]{wang20}
{Wang}, Y., {Beuther}, H., {Rugel}, M.~R., {et~al.} 2020, \aap, 634, A83

\bibitem[{{Wardle} \& {Yusef-Zadeh}(2002)}]{wardle02}
{Wardle}, M., \& {Yusef-Zadeh}, F. 2002, Science, 296, 2350

\bibitem[{{Wenger} {et~al.}(2018){Wenger}, {Balser}, {Anderson}, \&
  {Bania}}]{wenger18}
{Wenger}, T.~V., {Balser}, D.~S., {Anderson}, L.~D., \& {Bania}, T.~M. 2018,
  \apj, 856, 52

\bibitem[{{West} {et~al.}(2016){West}, {Safi-Harb}, {Jaffe}, {Kothes},
  {Landecker}, \& {Foster}}]{west16}
{West}, J.~L., {Safi-Harb}, S., {Jaffe}, T., {et~al.} 2016, \aap, 587, A148

\bibitem[{{Younes} {et~al.}(2017){Younes}, {Kouveliotou}, {Jaodand}, {Baring},
  {van der Horst}, {Harding}, {Hessels}, {Gehrels}, {Gill}, {Huppenkothen},
  {Granot}, {G{\"o}{\u{g}}{\"u}{\textcommabelow s}}, \& {Lin}}]{younes17}
{Younes}, G., {Kouveliotou}, C., {Jaodand}, A., {et~al.} 2017, \apj, 847, 85

\bibitem[{{Zhang} {et~al.}(2018){Zhang}, {Tian}, \& {Wu}}]{zhang18}
{Zhang}, M.~F., {Tian}, W.~W., \& {Wu}, D. 2018, \apj, 867, 61

\bibitem[{{Zhong} {et~al.}(2020){Zhong}, {Dai}, {Zhang}, \& {Deng}}]{zhong20}
{Zhong}, S.-Q., {Dai}, Z.-G., {Zhang}, H.-M., \& {Deng}, C.-M. 2020, \apjl,
  898, L5

\bibitem[{{Zhou} {et~al.}(2014){Zhou}, {Safi-Harb}, {Chen}, {Zhang}, {Jiang},
  \& {Ferrand}}]{zhou14}
{Zhou}, P., {Safi-Harb}, S., {Chen}, Y., {et~al.} 2014, \apj, 791, 87

\bibitem[{{Zhou} {et~al.}(2019){Zhou}, {Vink}, {Safi-Harb}, \&
  {Miceli}}]{zhou19}
{Zhou}, P., {Vink}, J., {Safi-Harb}, S., \& {Miceli}, M. 2019, \aap, 629, A51

\bibitem[{{Zhou} {et~al.}(2016){Zhou}, {Yang}, {Fang}, {Su}, {Sun}, \&
  {Chen}}]{zhou16c}
{Zhou}, X., {Yang}, J., {Fang}, M., {et~al.} 2016, \apj, 833, 4

\end{thebibliography}
\end{document}